\documentclass[twocolumn,prstab,floatfix]{revtex4}
\usepackage{graphicx}
\usepackage{amsmath}
\usepackage{latexsym}
\usepackage{amsfonts}
\usepackage{amssymb}

\begin{document}

\title{Optimal Axes of Siberian Snakes for Polarized Proton Acceleration}
\author{Georg H.~Hoffstaetter} 
\affiliation{Department of Physics, Cornell University, Ithaca/NY}

\begin{abstract}
  Accelerating polarized proton beams and storing them for many turns
  can lead to a loss of polarization when accelerating through
  energies where a spin rotation frequency is in resonance with orbit
  oscillation frequencies. First-order resonance effects can be avoided
  by installing Siberian Snakes in the ring, devices which rotate the
  spin by $180^\circ$ around the snake axis while not changing the
  beam's orbit significantly. For large rings, several Siberian Snakes
  are required.

  Here a criterion will be derived that allows to find an optimal
  choice of the snake axes.  Rings with super-period four are
  analyzed in detail, and the HERA proton ring is used as an example
  for approximate four-fold symmetry.  The proposed arrangement of
  Siberian Snakes matches their effects so that all spin-orbit
  coupling integrals vanish at all energies and therefore there is no
  first-order spin-orbit coupling at all for this choice, which I
  call snakes matching. It will be shown that in general at least eight
  Siberian Snakes are needed and that there are exactly four
  possibilities to arrange their axes.  When the betatron phase
  advance between snakes is chosen suitably, four Siberian Snakes can
  be sufficient.

  Since the spin motion depends on a particle's trajectory, protons at
  different phase space positions generally have different spin
  directions. The time averaged polarization at each phase space point
  is parallel to the invariant spin field, and the spread of this spin
  field limits the polarization that can be stored.  By the here
  presented choice of Siberian Snakes this limit is completely
  eliminated up to first order in the transverse phase space
  coordinates.  The invariant spin field and the amplitude dependent
  spin tune are also computed without linearization to show the
  advantages of this choice of snakes. Ultimately, the goal of snake
  matching is to reduce the loss of polarization during the
  acceleration of the beam. To show that favorable choice of snakes
  have been found, polarized protons are tracked for part of HERA-p's
  acceleration cycle which shows that polarization is preserved best
  for the here proposed arrangement of Siberian Snakes.
\end{abstract}
%\pacs{29.27.Hj,29.27.-a,29.20.-c,41.75.Lx}

\maketitle

\section{Introduction}

The design orbit spin direction $\vec n_0$ at some azimuth
$\theta=2\pi l/L$ along a storage ring of length $L$ describes the
spin direction for the design orbit which is periodic from turn to
turn, where $l$ is the pathlength along the design orbit.  The design
orbit spin tune $\nu_0$ describes the number of times a spin has
rotated around $\vec n_0(\theta)$ during one turn around the ring.

Spin-orbit resonances in high energy accelerators arise when the
electro-magnetic fields on synchro-betatron trajectories cause
disturbances of the spin's motion which build up coherently from turn
to turn.

In a flat ring, for instance, an initially vertical spin of a particle
traveling on the design orbit remains vertical during particle motion.
On a vertical betatron trajectory the particle traverses horizontal
fields in quadrupoles and the spin no longer remains vertical.  This
disturbance of spin motion due to the betatron motion is described by
the spin-orbit coupling integrals~\cite{steffen88a} 
\begin{equation}
I^\pm=\int_0^{2\pi}k_y{\rm e}^{{\rm i}(-\Psi\pm\Phi_y)}d\theta\ ,
\label{eq:socoup}
\end{equation}
where $k_y=k\sqrt{\beta_y}$ with the quadrupole strength $k$ and the
vertical beta function $\beta_y$, and $\Psi$ is the phase advance of
the spin rotation around the vertical and $\Phi_y$ is the vertical
betatron phase.  For ease of notation a constant factor which is
sometimes multiplied to this integral is not indicated here.

The $I^\pm$ tend to be especially big when the spin disturbance in
every FODO cell of a ring adds up coherently~\cite{courant85}.

In an approximation that is of first order in transverse phase space
coordinates, these integrals yield the following important
information: If all spin-orbit coupling integrals vanish, all
initially vertical spins are again vertical after one turn, although
they have traveled along different betatron trajectories.  The ring is
then called spin matched or spin transparent.

In this paper it will be analyzed how Siberian Snakes can be used to
make a storage ring spin transparent at all energies.

When a beam is polarized, it has a polarization direction $\vec f(\vec
z,j)$ at each phase space position $\vec z$ at some azimuth $\theta$
of the ring, where $j$ indicates how often the bunch has traveled
around the ring. Since each initial phase space position leads to a
different particle trajectory with different spin motion, spin fields
in general change from turn to turn. A special spin field that does
not change from turn to turn is called the invariant spin field $\vec
n(\vec z)$~\cite{hoffstaetter98e}.

When a beam is polarized according to this spin field, $\vec f(\vec
z,j)=\vec n(\vec z)$, then the beam polarization is given by the phase
space average $P_{lim}=<\vec n(\vec z)>_{\vec z}$.  In any case, even
when a particle at $\vec z(j)$ after its $j$th turns has a classical
spin vector $\vec S$ with a direction which is not parallel to $\vec
n(\vec z(j))$, its projection onto the invariant spin field is
constant, i.e. $\vec f(\vec z(j),j)\cdot\vec n(\vec z(j))$ does not
change with $j$. This is possible since its spin $\vec S$ appears to
rotate around $\vec n(\vec z(j))$. The time averaged polarization at
the phase space point $\vec z$ is therefore parallel to $\vec n(\vec
z)$ for any spin field. The maximum time averaged polarization at that
point is thus given when the beam is initially polarized parallel to
$\vec n(\vec z)$.  The average $P_{lim}$ is therefore called the
limiting polarization~\cite{hoffstaetter00, vogt00, hoffstaetter98c}.

To illustrate one of the benefits of the proposed choice of Siberian
Snakes it will be shown that they increase $P_{lim}$.

When an invariant spin field is found, the rotation of $\vec S$ around
it can be described by an amplitude dependent spin tune $\nu(\vec J)$
that depends on the orbital amplitudes $\vec J$ but not on the orbital
phase variables. This is important since $\nu(\vec J)$ does not change
from turn to turn and can therefore be used to describe long term
coherence with the frequencies of orbital motion. When this amplitude
dependent spin tune is in resonance with the orbital tunes
$Q_j$,i.e. $\nu(\vec J)=k_0+\sum_j k_j Q_j$ for integers $k_j$, the
spin motion can be strongly disturbed.

As another illustration of the benefits of the proposed choice of
Siberian Snakes it will be shown that they increase the orbital
amplitudes for which low order resonance conditions occur.

It can be shown that the projection of the polarization onto the
invariant spin field $J_S=\vec S\cdot\vec n(\vec z)$ is not only
invariant in a storage ring, but that it is an adiabatic invariant
\cite{hoffstaetter02a} when parameters of the accelerator, like the
beam's energy, change.  This means that a beam that is initially
polarized parallel to the invariant spin field $\vec n(\vec z,E_i)$
and thus has the average polarization $P_{lim}(E_i)$ at its initial
energy $E_i$ will still be polarized parallel to the invariant spin
field $\vec n(\vec z,E_f)$ after the beam has been accelerated to a
final energy $E_f$, and its average polarization is
$P_{lim}(E_f)$. This final polarization can be very large, even though
it may have been very small at some intermediate energies.  All this
is only true if the change of energy is performed adiabatically
slowly. Usually, the acceleration cannot be performed adiabatically
slowly at all energies and polarization is lost, i.e. the final
polarization is smaller than $P_{lim}(E_f)$.

The most convincing illustration of the benefits of the proposed
choice of Siberian Snakes is finally that they increase the
polarization that is retained after acceleration.

Although it has been straightforward to define ${\vec n}(\vec z)$, it
is not easy to calculate this spin field in general and much effort
has been spent on this topic, mostly for electrons at energies up to
46~GeV. All algorithms developed before the polarized proton project
at HERA-p rely on perturbation methods at some stage, and either do
not go to high enough order \cite{chao81b,eidelmann94a} or have
problems with convergence at high order and high proton energies
\cite{mane87b,yokoya92,balandin92}. The algorithms developed for the
HERA-p project
\cite{hoffstaetter96d,hoffstaetter99k,hoffstaetter00,vogt00,hoffstaetter02b}
made the here presented analysis possible.

\section{Optimal Choices of Siberian Snakes}
\label{sc:optsnake}

Siberian Snakes are indispensable if polarized proton beams are to be
accelerated in a high-energy synchrotron such as HERA-p.  This has
the following reasons:
\begin{enumerate}
\item Siberian Snakes fix the design-orbit spin tune $\nu_0$ to
  $\frac{1}{2}$ during the acceleration cycle so that no first-order
  resonances have to be crossed.  Crossing first-order resonances
  can lead to a severe reduction of polarization by an amount
  described by the Froissart-Stora formula~\cite{froissart60}.
\item Siberian Snakes strongly reduce the influence of energy
  variations on spin motion within a synchrotron period~\cite{balandin96a}.
\item Siberian Snakes reduce the variation of $\vec n(\vec z)$ for
  particles which oscillate vertically and therefore pass through
  horizontal fields which perturb the spin motion.
\item When $\vec n$ changes rapidly during acceleration, the adiabatic
  invariance of $J_S=\vec n(\vec z)\cdot\vec S$ might be violated and
  polarization would be reduced.  It is therefore important that
  Siberian Snakes smoothen the changes of $\vec n$ during the
  acceleration cycle.
\item Siberian Snakes can also compensate perturbing effects of
  misaligned optical elements \cite{derbenev76,balandin96a,lee97a} but
  the effect of misalignments will not be covered here.
\end{enumerate}

There is so far no reliable formula for determining the number of
Siberian Snakes required for an accelerator \cite{anferov99b,chao99a}.
To make things worse, for any given number of Siberian Snakes there
are very many different possible combinations of the snake angles
which lead to an energy independent closed-orbit spin tune of
$\frac{1}{2}$ and to a vertical design orbit spin direction $\vec n_0$
in the accelerator's arcs. But so far there has been no reliable
formula for determining which of these snake schemes leads to the
highest polarization.

There used to be a popular opinion that, owing to their symmetry, 5
standard choices of the snake angles for 4 Siberian Snakes are
advantageous for HERA-p.  These choices are not optimal, as will be
shown.  For reasons why these standard schemes were considered useful
see for example \cite{ptitsin96b}.  RHIC with its two snakes, is
operated with a similar standard scheme~\cite{courant95a}.
\begin{figure}[ht!]
\begin{center}
\begin{minipage}[t]{\columnwidth}
\includegraphics[width=\columnwidth, bb=124 524 548 763, clip]
                {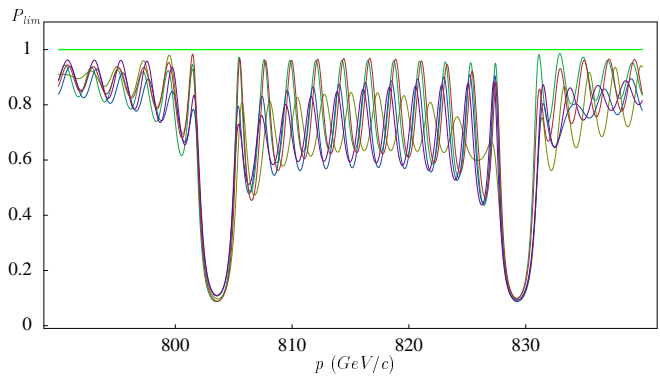}
\end{minipage}
\end{center}
\caption[$P_{lim}$ of linearized spin-orbit motion for 5 standard
         snake schemes] {$P_{lim}$ of linearized spin-orbit motion
         for a 2.5$\sigma$ vertical amplitude for 4 Siberian Snakes in
         HERA-p.  Each curve corresponds to one of the 5 standard
         choices of snake schemes in the following table.  The right
         column shows an obvious notation to describe the snake angles
         in a snake scheme:\\
\begin{minipage}{\columnwidth}
\begin{center}
\begin{tabular}{lrrrr}
Scheme & South        & East          & North        & West         \\
\hline
$(0\frac{\pi}{2}00)$&
       $    0^\circ$ &$   90^\circ$ &$    0^\circ$ &$    0^\circ$ \\
$(\frac{\pi}{4}0\frac{\pi}{4}0)$&
       $   45^\circ$ &$    0^\circ$ &$   45^\circ$ &$    0^\circ$ \\
$(\frac{3\pi}{4}0\frac{3\pi}{4}0)$&
       $  -45^\circ$ &$    0^\circ$ &$  -45^\circ$ &$    0^\circ$ \\
$(\frac{5\pi}{8}\frac{\pi}{8}\frac{5\pi}{8}\frac{\pi}{8})$&
       $-22.5^\circ$ &$ 22.5^\circ$ &$-22.5^\circ$ &$ 22.5^\circ$ \\
$(\frac{\pi}{8}\frac{5\pi}{8}\frac{\pi}{8}\frac{5\pi}{8})$&
       $ 22.5^\circ$ &$-22.5^\circ$ &$ 22.5^\circ$ &$-22.5^\circ$ \\
\end{tabular}
\end{center}
\end{minipage}
}
\label{fg:slimherast}
\end{figure}

\begin{figure}[ht!]
\begin{center}
\begin{minipage}[t]{\columnwidth}
\includegraphics[width=0.49\columnwidth,bb=117 629 336 755,clip]
                {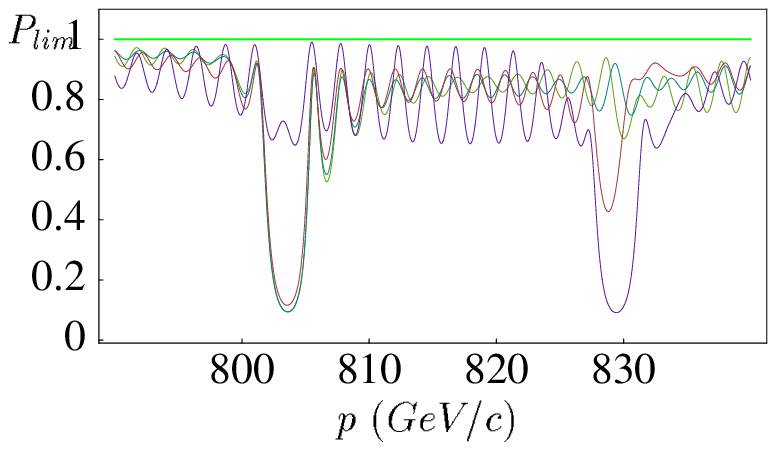}
\includegraphics[width=0.49\columnwidth,bb=113 629 337 761,clip]
                {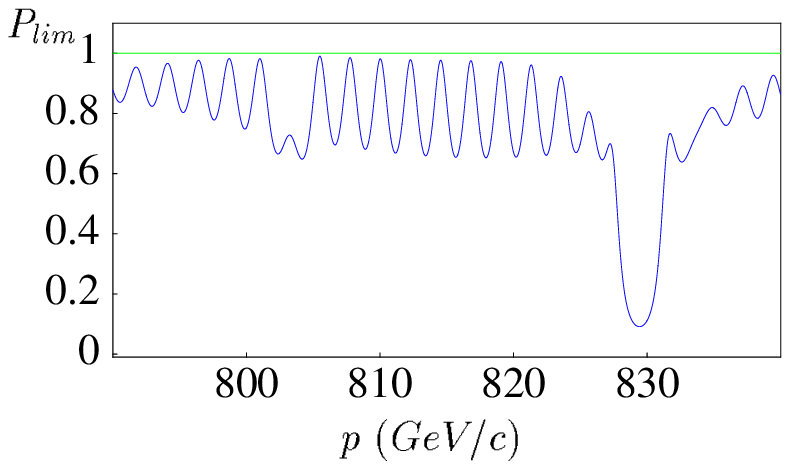}
\end{minipage}
\end{center}
\caption[{\bf Left}: Snake schemes with very different $P_{lim}$]
 {$P_{lim}$ for 4 different snake schemes which lead to very different
 maximum time average polarization for the standard HERA-p
 lattice. {\bf Right}: One of these snake schemes which were found by
 numerical optimization of the snake angles.
\label{fg:schemex}}
\end{figure}

The energy dependence of $P_{lim}$ in HERA-p produced by these 5 snake
schemes is shown in Fig.~\ref{fg:slimherast}.  They seem to produce
rather similar but very low maximum time average polarization
$P_{lim}$ in a critical energy regions where very strong resonances
are excited.  The observation of such rather small differences in the
$\vec n$-axis for such different schemes suggests the following
detailed investigation of the influence of snake schemes.  Figure
\ref{fg:schemex}~(left) shows $P_{lim}$ as computed for linearized
spin-orbit motion \cite{hoffstaetter00} for 4 other schemes with 4
Siberian Snakes which were found by the so called filtering method
\cite{hoffstaetter00,hoffstaetter96e}, a numerical search for suitable
snake axes.  It is apparent that large increases in $P_{lim}$ can
result from the choice of a suitable snake scheme.

HERA-p is the proton ring of an electron-proton collider and is
located above the electron ring. It therefore has non-flat sections in
which the protons are bend down to collide with the electrons. These
sections bend spins out of the vertical so that $\vec n_0$ would be
non-vertical in the arcs. To avoid this, it has been proposed to
insert a Siberian Snake in each of the 6 vertically bending
regions. These so called flattening snakes have been assumed in all
presented computations in addition to the Siberian Snakes for which an
optimal axis is sought. As a 1$\sigma$ emittance a realistic value of
4$\pi$mm~mrad was assumed throughout this paper.

\section{Spin-Orbit-Coupling Integrals}

For the spin-orbit-coupling integrals in flat rings of
Eq.~(\ref{eq:socoup}), $\vec n_0$ was assumed to point vertically
upward. Now Siberian Snakes will be included, which rotate all spins
and also $\vec n_0$ by $\pi$ around some axis in the horizontal plane,
so that $\vec n_0$ points downward in some sections. This changes the
sign of $\Psi$ Eq.~(\ref{eq:socoup}).

It is assumed that there are $n$ Siberian Snakes in the ring and that
$n$ is even, to make $\vec n_0$ vertical in the arcs of the ring.  The
azimuth at the position of a snake is denoted by $\theta_j$ and the
spin phase advance around the vertically upward direction between
snake $j$ and $j+1$ is denoted by $\Psi_j$.  The spin phase advance
after the $j$th Siberian Snake is $\Psi_j(\theta)$ with
$\Psi_j(\theta_j)=0$.

For simplicity $\theta_0=0$ and $\theta_{n+1}=2\pi$ is used and
the spin phase advance from azimuth $\theta_0$ to the first
Siberian Snake is $\Psi_0$.

In the following a Siberian Snake with a snake axis which is in the
horizontal plane will be referred to as a horizontal Siberian Snake
and for historical reasons a snake which rotates spins around the
vertical by some rotation angle will be referred to as type~III snake.
A horizontal Siberian Snake with a snake angle $\varphi$ is equivalent
to a radial Siberian Snake followed by a type~III snake that introduces
an extra spin phase advance of $\alpha=2\varphi$~\cite{montague84}.

With these notations the spin-orbit-coupling integrals for a ring with
horizontal Siberian Snakes are
\begin{eqnarray}
I_y^\pm &=& -\sum_{j=0}^n
\text{e}^{-\text{i}\sum\limits_{k=0}^{j-1}(-)^k(\alpha_k+\Psi_k)}
\nonumber\\ &\times&
\int\limits_{\theta_j}^{\theta_{j+1}}k_y
\text{e}^{\text{i}[-(-)^j(\Psi_j(\theta)+\alpha_j)\pm\Phi]}\text{d}\theta\;
.
\end{eqnarray}
In terms of the orbital phase advance $\Phi_{j}$ between snake $j$
and $j+1$, one obtains
\begin{eqnarray}
I_y^\pm &=& -\sum_{j=0}^n
\text{e}^{\text{i}\sum\limits_{k=0}^{j-1}[-(-)^k(\alpha_k+\Psi_k)\pm\Phi_{k}]}
\nonumber\\ &&
\times\int\limits_{\theta_j}^{\theta_{j+1}}k_y
\text{e}^{\text{i}[-(-)^j(\Psi_j(\theta)+\alpha_j)\pm\Phi_{j}(\theta)]}
\text{d}\theta\; .
\label{eq:socoupss}
\end{eqnarray}

A corresponding formula has been used in \cite{steffen88a} to introduce
so-called strong spin matching, where Siberian Snakes are used to
produce a cancelation of spin perturbations in different FODO cells.

The spin phase advances between snakes must satisfy the condition
$\sum_{k=0}^n(-)^k\Psi_k=0$ to make the closed-orbit spin tune
independent of energy and the snake angles must satisfy the condition
$\sum_{k=1}^n(-)^k\varphi_k=\frac{\pi}{2}$ to make the closed-orbit
spin tune $\nu_0$ equal to $\frac{1}{2}$.

\section{Snake Matching in Rings with Super-Periodicity}
\label{sc:snmatch}

The spin perturbations in different parts of the ring can compensate
each other when these parts have similar spin-orbit-coupling
integrals.  This is achieved by using the Siberian Snakes to adjust
the spin phase advances in such a way that spin-orbit-coupling
integrals of similar parts of a ring cancel each other.  In the
following, the process of finding a snake scheme for which such a
compensation occurs will be referred to as \emph{snake matching}
\cite{hoffstaetter03,hoffstaetter99b2}.  After demonstrating the idea
for type~III snakes, which simply rotate spins around the vertical by
some fixed angle with little influence on the orbit motion, two quite
general results will be demonstrated:
\begin{enumerate}
\item A ring with super-periodicity 4 can be completely snake matched
  using 8 Siberian Snakes, i.e. a snake scheme can be found for which
  the spin-orbit-coupling integrals are zero due to a complete
  cancelation of spin perturbations in different parts of the ring.
  There are exactly two such possibilities which lead to energy
  independent snake angles.
\item Such a ring can also be snake matched using 4 Siberian Snakes.
  Then, however, the snake axes depend on energy and have to be
  changed during the acceleration process.
\item Energy independent snake matching with 4 Siberian Snakes can be
  found when the betatron phase advance is appropriately chosen for
  each of the 4 quadrants.
\end{enumerate}

\subsection{Snake Matching with Type~III Snakes for Super-Periodicity 4:}

The index $y$ on the spin-orbit-coupling integral and on the vertical
phase advance and tune will not be indicated.  In any case, the
methods for canceling spin-orbit-coupling integrals by a special
choice of snake angles which will now be derived can also be used for
transverse and longitudinal motion.  In this section the notation will
be further simplified by using the symbols $\nu_0$ and $Q$ to denote
$2\pi$ times the spin tune and $2\pi$ times the orbital tune.  Then for a
ring with super-periodicity 4, $I^\pm$ can be computed from
\begin{eqnarray}
I^\pm_\frac{1}{4}&=&\int_0^{\pi/2}
k_y \text{e}^{\text{i}(-\Psi\pm\Phi)}\text{d}\theta\; ,
\label{eq:1234}\\
I^\pm&=&I^\pm_{\frac{1}{4}}[1+ \text{e}^{\text{i}(-\nu_0\pm Q)/4}
+ \text{e}^{\text{i}2(-\nu_0\pm Q)/4}
+ \text{e}^{\text{i}3(-\nu_0\pm Q)/4}]\; .\nonumber
\end{eqnarray}
Spin transparency requires that $I^+$ as well as $I^-$ vanish.  Thus
the bracket in (\ref{eq:1234}) must vanish. This is only possible when
$\text{e}^{\text{i}(-\nu_0\pm Q)/4}$ is either $-1$ or $i$.  Choosing
the first possibility to eliminate $I^+$ and the second to eliminate
$I^-$, one obtains
\begin{equation}
\text{e}^{\text{i}(-\nu_0+ Q)/4}=-1\; ,\ \ \text{e}^{\text{i}(-\nu_0-
Q)/4}=i\; .
\end{equation}
This leads to the requirement
$\text{e}^{\text{i} Q/2}=i$ which cannot be satisfied in a realistic
ring.  Therefore, a four-fold repetitive symmetry cannot lead to spin
transparency at any energy.
While the spin disturbance of two quadrants can therefore not cancel
in $I^+$ as well as in $I^-$, one of these integrals can cancel
whenever the spin phase advance between the quadrants is appropriate.

The situation changes if type~III snakes are installed.  As first
found in \cite{hoffstaetter96e}, type~III snakes can improve the spin
dynamics in HERA-p by increasing $P_{lim}=$\mbox{$|\langle\vec
n\rangle|$}.  They can be used to manipulate the spin phase advance to
make the spin-orbit-coupling integrals of different parts of the ring
cancel.  To demonstrate this, 4 type~III snakes are installed
regularly spaced around the ring.

\begin{figure}[ht!]
\begin{center}
\begin{minipage}[t]{\columnwidth}
\includegraphics[width=0.32\columnwidth,bb=123 629 265 756,clip]
                {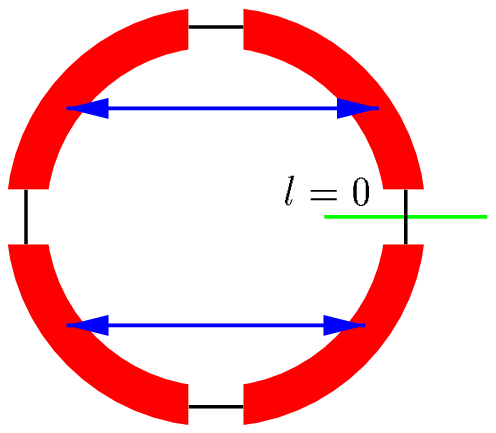}
\includegraphics[width=0.32\columnwidth,bb=123 629 265 756,clip]
                {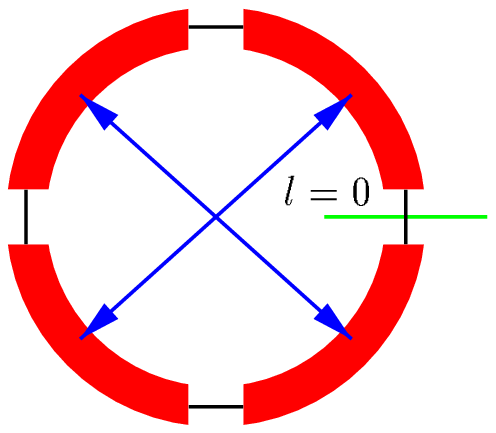}
\includegraphics[width=0.32\columnwidth,bb=123 632 265 755,clip]
                {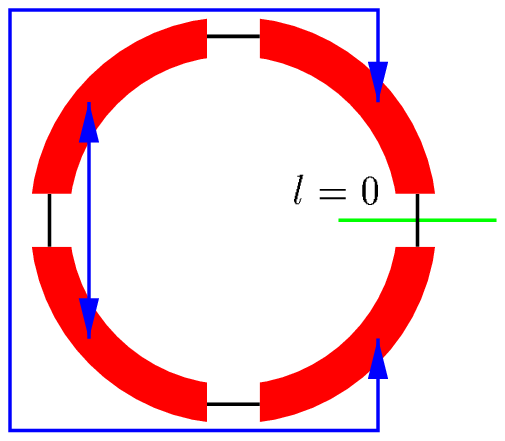}
\end{minipage}
\end{center}
\caption[Three possibilities for canceling the depolarizing effects of
  quadrants]
 {The three possibilities for canceling the depolarizing effects of
  quadrants of a ring with super-periodicity 4.  The arrows indicate
  which quadrants cancel
\label{fg:cancel}}
\end{figure}

There are three possibilities for canceling the spin disturbances
between quadrants of the ring.  The quadrants whose destructive
effects cancel are connected by arrows in Fig.~\ref{fg:cancel}.

\begin{figure}[ht!]
\begin{center}
\begin{minipage}[t]{0.5\columnwidth}
\includegraphics[width=\columnwidth,bb=117 554 338 757,clip]
                {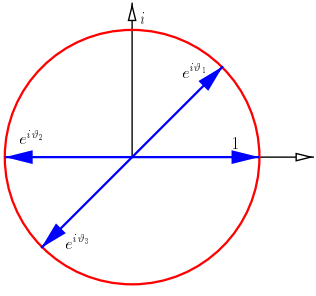}
\end{minipage}
\end{center}
\caption[Four ways for complex numbers to add up to zero]
 {Four complex numbers with modulus one
  can only add up to zero when they consist of two pairs which
  individually add up to zero
\label{fg:4c}}
\end{figure}

The spin-orbit-coupling integrals are
\begin{eqnarray}
I^\pm
&=&
I^\pm_{\frac{1}{4}}[1+ \text{e}^{\text{i}(-\nu_0\pm Q)/4-\psi_1}
\label{eq:4c}\\
&& +
 \text{e}^{\text{i}2(-\nu_0\pm Q)/4-\psi_1-\psi_2}
+ \text{e}^{\text{i}3(-\nu_0\pm Q)/4-\psi_1-\psi_2-\psi_3}]\; ,\nonumber
\end{eqnarray}
where $\psi_j$ is the spin rotation angle of the type~III
snake at $\theta=j\frac{\pi}{2}$.
To snake match the ring, $I^+$ as well as $I^-$ must vanish.
Therefore the bracket on the right hand side has to vanish in both
cases.  A sum of 4 complex numbers with unit modulus can only vanish
when it consist of two pairs of numbers which cancel each other.  This
is shown in Fig.~\ref{fg:4c}.

The three possibilities of cancelation demonstrated in
Fig.~\ref{fg:cancel} are given by the following three sets of
equations:
\begin{equation}
\begin{array}{lccc}
1.        & (-\nu_0\pm Q)/4-\psi_1 & \stackrel{\circ}= & \pi\\
\text{and}& (-\nu_0\pm Q)/4-\psi_3 & \stackrel{\circ}= & \pi\\
2.        & 2(-\nu_0\pm Q)/4-\psi_1-\psi_2 & \stackrel{\circ}= & \pi\\
\text{and}& \psi_3 & \stackrel{\circ}= & \psi_1\\
3.        & 3(-\nu_0\pm Q)/4-\psi_1-\psi_2-\psi_3 & \stackrel{\circ}= & \pi\\
\text{and}& (-\nu_0\pm Q)/4-\psi_2 & \stackrel{\circ}= & \pi\\
\end{array}
\end{equation}
The symbol $\stackrel{\circ}=$ indicates equivalence modulo $2\pi$.
To snake match, one of these three conditions has to hold for
$(-\nu_0+Q)$, which lets $I^+$ vanish and another of the conditions
has to hold for $(-\nu_0-Q)$, which lets $I^-$ vanish.  $I^+$ and
$I^-$ cannot vanish due to the same condition if restrictions on the
allowed orbital phase advance $Q$ are to be avoided.  There are
therefore three possibilities:
\begin{enumerate}
\item $I^+=0$ due to condition 2 and $I^-=0$ due to condition 3 requires
\begin{eqnarray}
\psi_2&\stackrel{\circ}=&\pi+2(-\nu_0+Q)/4-\psi_1\; ,\ \
\psi_3\stackrel{\circ}=\psi_1\; ,\\
\psi_2&\stackrel{\circ}=&\pi+(-\nu_0-Q)/4\; ,\ \
\psi_3\stackrel{\circ}=2(-\nu_0-Q)/4-\psi_1\; .
\nonumber
\end{eqnarray}
The first and the third of these equations require that
$\psi_1\stackrel{\circ}=(-\nu_0+3Q)/4$ whereas the second and the
fourth equations require that
$\psi_1\stackrel{\circ}=(-\nu_0-Q)/4$. These two requirements are in
general not compatible and the ring cannot be made spin transparent in
this way.
\item $I^+=0$ due to condition 1 and $I^-=0$ due to condition 3 requires
\begin{eqnarray}
\psi_1&\stackrel{\circ}=&\pi+(-\nu_0+Q)/4\; ,\ \
\psi_3\stackrel{\circ}=\psi_1\; ,\\
\psi_2&\stackrel{\circ}=&\pi+(-\nu_0-Q)/4\; ,\ \
\psi_3\stackrel{\circ}=2(-\nu_0-Q)/4-\psi_1\; .
\nonumber
\end{eqnarray}
The first and the last of these equations together require
$\psi_3\stackrel{\circ}=\pi-(3Q-\nu_0)/4$.  This is in conflict with
the second equation.  Thus this way also cannot lead to a spin
transparent ring.
\item $I^+=0$ due to condition 1 and $I^-=0$ due to condition 2 requires
\begin{eqnarray}
\psi_1&\stackrel{\circ}=&\pi+(-\nu_0+Q)/4\; ,\ \
\psi_3\stackrel{\circ}=\psi_1\; ,\\
\psi_2&\stackrel{\circ}=&\pi+2(-\nu_0-Q)/4-\psi_1\; ,\ \
\psi_3\stackrel{\circ}=\psi_1\; .
\nonumber
\end{eqnarray}
These 4 equations are compatible and lead to
$\psi_1\stackrel{\circ}=\psi_3\stackrel{\circ}=\pi+(-\nu_0+Q)/4$ and
$\psi_2\stackrel{\circ}=(-\nu_0-3Q)/4$.
\end{enumerate}

The type~III snake at $l=0$ has the rotation angle $\psi_4$ which is
chosen in such a way that the closed-orbit spin tune of the ring
does not change due to the snakes,
i.e.~$\psi_1+\psi_2+\psi_3+\psi_4\stackrel{\circ}=0$.  The required
rotation angles are then
\begin{eqnarray}
\psi_1&\stackrel{\circ}=&\psi_3\stackrel{\circ}=\pi+\frac{-\nu_0+Q}{4}\; ,\\
\psi_2&\stackrel{\circ}=&-\frac{\nu_0+3Q}{4}\; ,\ \
\psi_4\stackrel{\circ}= \frac{Q-3-\nu_0}{4}\; .
\nonumber
\label{eq:4ca}
\end{eqnarray}

\begin{figure}[ht!]
\begin{center}
\begin{minipage}[t]{\columnwidth}
\includegraphics[width=0.49\columnwidth,bb=119 538 344 755,clip]
                {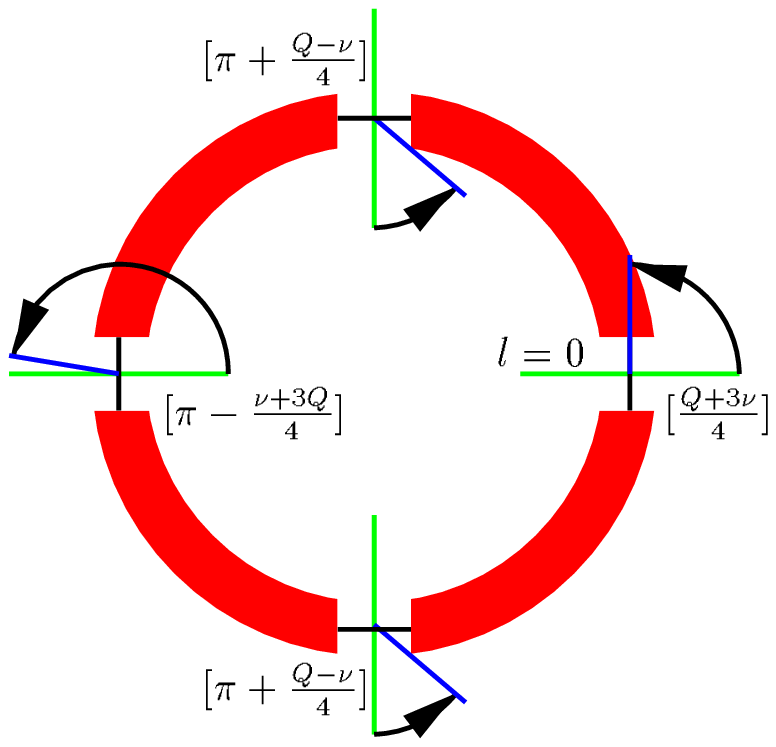}
\includegraphics[width=0.49\columnwidth,bb=119 538 344 755,clip]
                {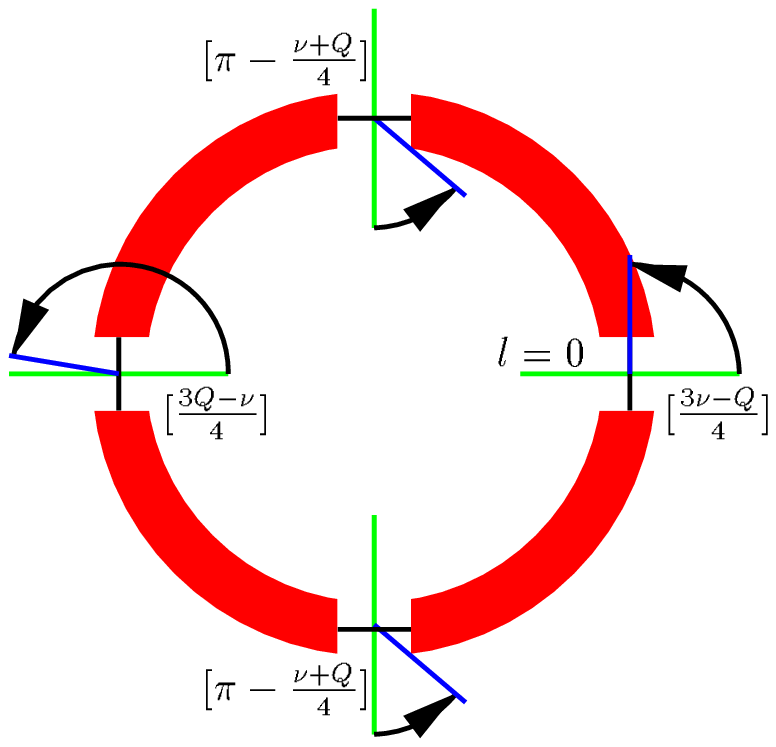}
\end{minipage}
\end{center}
\caption[Two ways to snake match with 4 type~III snakes]
 {The only two ways to snake match a ring with super-periodicity 4
  by 4 type~III snakes. The number $[x]$ denotes $x$ mod $2\pi$ and lies
  in $[0,2\pi)$.  The vertical tune
  times $2\pi$ is denoted by $Q$ and $\nu_0=G\gamma 2\pi$
\label{fg:4t3}}
\end{figure}

Obviously a change in sign of $Q$ leads to $I^+=0$ due to condition 2
and to $I^-=0$ due to condition 1.  There are therefore exactly two
possibilities for making a ring with super-periodicity 4 spin
transparent by means of 4 type~III snakes.  These possibilities are
shown in Fig.~\ref{fg:4t3}.
However, the scheme of 4 type~III snakes presented here cannot be a
practical snake scheme, since it does not make the closed-orbit spin
tune independent of energy.  But it illustrates
how type~III snakes can be used at fixed energy to make spin
perturbations from different parts of the ring cancel each other. This
feature can then be used in combination with the Siberian Snakes which
are installed to make the closed-orbit spin tune independent of
energy.

\subsection{Snake Matching with Type~III Snakes for Super-Periodicity 4
  and Mirror Symmetry:}

In particle optical systems, mirror symmetries are often used to
cancel perturbative effects
\cite{rose87,degenhardt92,wan95,hoffstaetter98f}.  Therefore it is
interesting to see whether mirror symmetry can lead to vanishing
of spin-orbit-coupling integrals when 4 Siberian Snakes are
installed at the symmetry points of the ring. If one super-period is
mirror symmetric, then
\begin{eqnarray}
I^\pm_\frac{1}{8}
&=&
\int_0^{\pi/4}
k_y \text{e}^{\text{i}(-\Psi\pm\Phi)}\text{d}\theta\; ,\\
I^\pm_\frac{1}{4}
&=&
I^\pm_\frac{1}{8}
+(I^\pm_\frac{1}{8})^* \text{e}^{\text{i}(-\nu_0\pm Q)/4}\; .
\label{eq:mir}
\end{eqnarray}
With $\Sigma^\pm=(-\nu_0\pm Q)/4$ one obtains for the complete ring
\begin{eqnarray}
I^\pm=(I^\pm_\frac{1}{8}+(I^\pm_\frac{1}{8})^* \text{e}^{\text{i}\Sigma^\pm})
\cdot
(1+ \text{e}^{\text{i}\Sigma^\pm}
+ \text{e}^{\text{i}2\Sigma^\pm}
+ \text{e}^{\text{i}3\Sigma^\pm}).
\label{eq:8cond}
\end{eqnarray}

Type~III snakes will not be considered further, since a
horizontal Siberian Snake with snake angle $\varphi$ can be decomposed
into a radial Siberian Snake and a type~III snake with rotation angle
$2\varphi$.

\subsection{Snake Matching of Siberian Snakes with Fixed Axes at All
Energies for Super-Periodicity 4:}
\label{sc:match4}

Thus snake matching the ring with super-periodicity 4 is not
influenced by the fact that the ring might have a mirror symmetry
since the bracket in (\ref{eq:8cond}) is equivalent to the
corresponding bracket in (\ref{eq:1234}) for rings without
mirror symmetry.

\paragraph{Schemes with 4 Snakes:}
For 4 horizontal Siberian Snakes the spin-orbit-coupling integral in
(\ref{eq:socoupss}) is
\begin{eqnarray}
I^\pm
&=&
\int_{\theta_0}^{\theta_1}k_y
 \text{e}^{\text{i}(-\Psi_0\pm\Phi_{0})}\text{d}\theta\\
& &+\text{e}^{\text{i}(-\Psi_0\pm\Phi_{0})}
\int_{\theta_1}^{\theta_2}k_y
 \text{e}^{\text{i}(\Psi_1+\alpha_1\pm\Phi_{1})}\text{d}\theta\nonumber\\
& &+\text{e}^{\text{i}(-\Psi_0+\alpha_1+\Psi_1\pm(\Phi_{0}+\Phi_{1}))}
\nonumber\\
&\times&
\int_{\theta_2}^{\theta_3}k_y
 \text{e}^{\text{i}(-\Psi_2-\alpha_2\pm\Phi_{2})}\text{d}\theta\nonumber\\
& &+\text{e}^{\text{i}(-\Psi_0+\alpha_1+\Psi_1-\alpha_2-\Psi_2
\pm(\Phi_{0}+\Phi_{1}+\Phi_{2}))}
\nonumber\\
&\times&
\int_{\theta_3}^{2\pi}k_y
 \text{e}^{\text{i}(\Psi_3+\alpha_3\pm\Phi_{3})}\text{d}\theta\; .\nonumber
\end{eqnarray}
For a ring with super-periodicity 4 and with 4 equally spaced horizontal
Siberian Snakes one obtains with $\nu_0=\Psi(2\pi)$,
$\Psi_j=\Psi(2\pi)/4$, and $\Phi_{j}=Q/4$ the relation
\begin{eqnarray}
I^\pm_{\frac{1}{4}}&=& \int_{0}^{\frac{\pi}{2}}k_y
\text{e}^{\text{i}(-\Psi\pm\Phi)}\text{d}\theta\; ,\\ I^\pm
&=&I^\pm_{\frac{1}{4}}( 1+ \text{e}^{\text{i}(-\Psi_0+\alpha_1 +
\Psi_1-\alpha_2\pm(\Phi_{0}+\Phi_{1}))})\\ &
&+(I^\mp_{\frac{1}{4}})^*
\text{e}^{\text{i}(-\Psi_0+\alpha_1\pm\Phi_{0})}
\nonumber\\
&\times&
(1+
\text{e}^{\text{i}(\Psi_1-\alpha_2-\Psi_2+\alpha_3\pm(\Phi_{1}+\Phi_{2}))})
\nonumber\\ &=&I^\pm_{\frac{1}{4}}(1+
\text{e}^{\text{i}(\alpha_1-\alpha_2\pm2Q/4)})\nonumber\\ && +
(I^\mp_{\frac{1}{4}})^* \text{e}^{\text{i}((-\nu_0\pm Q)/4+\alpha_1)}
(1+ \text{e}^{\text{i}(-\alpha_2+\alpha_3\pm2Q/4)})\; .\nonumber
\end{eqnarray}
Spin transparency of the ring is therefore obtained when
\begin{equation}
\alpha_1-\alpha_2\pm 2Q/4\stackrel{\circ}=\pi\ \mathrm{and\ }
-\alpha_2+\alpha_3\pm 2Q/4\stackrel{\circ}=\pi\; .
\label{eq:no4sol}
\end{equation}
This cannot be achieved in general since the conditions
$\alpha_1-\alpha_2+2Q/4\stackrel{\circ}=\pi$ and
$\alpha_1-\alpha_2-2Q/4\stackrel{\circ}=\pi$ have to be satisfied
simultaneously, which implies $Q\stackrel{\circ}=0$.

\begin{figure}[ht!]
\begin{center}
\begin{minipage}[t]{\columnwidth}
\includegraphics[width=0.49\columnwidth,bb=121 530 336 755,clip]
                {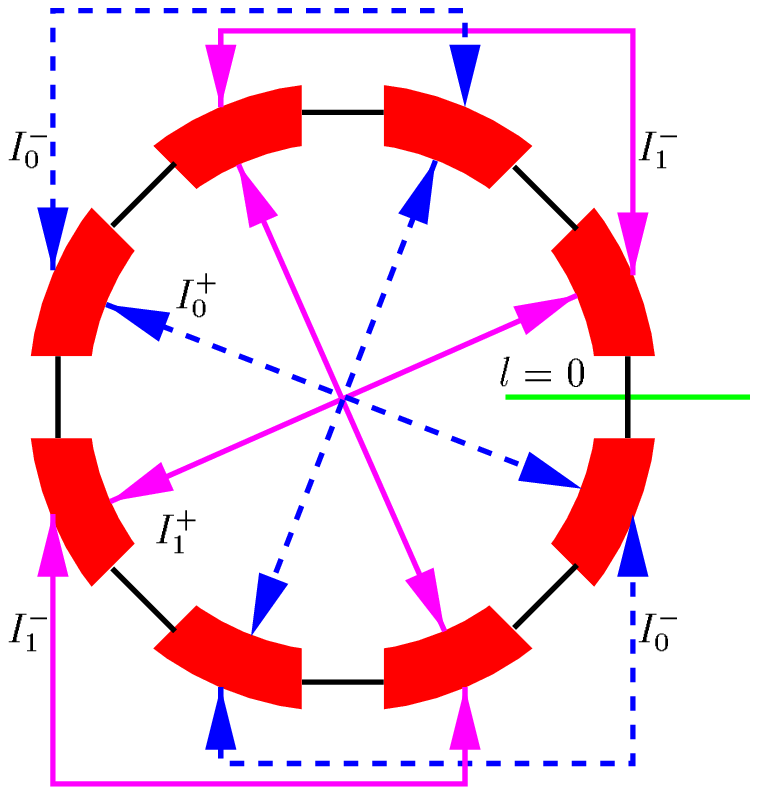}
\includegraphics[width=0.49\columnwidth,bb=121 530 336 755,clip]
                {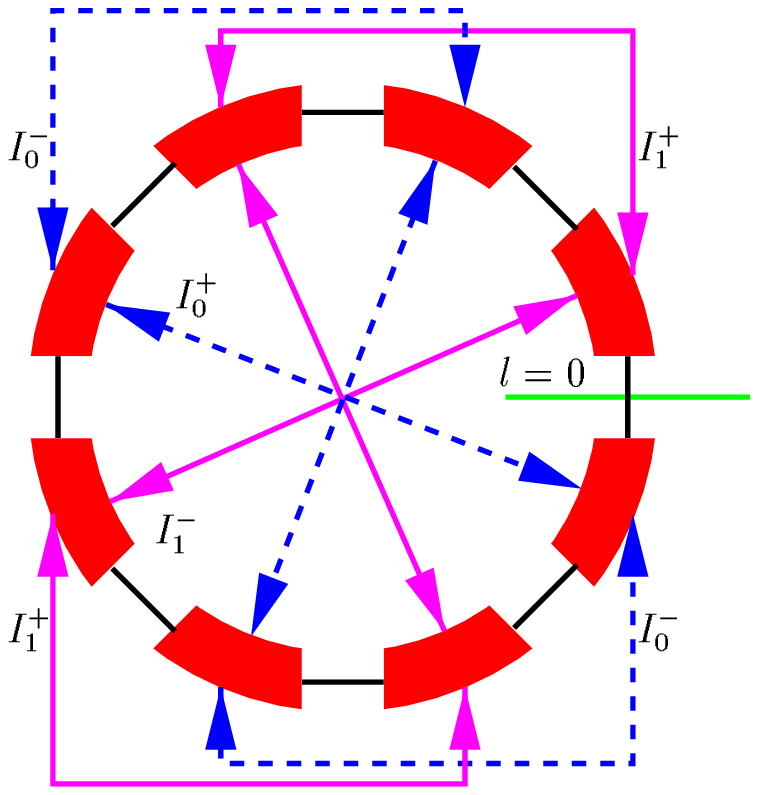}
\end{minipage}
\end{center}
\caption[The two possibilities for spin perturbations of octants to cancel]
 {The two possibilities by which the individual parts of the
  ring cancel the spin-orbit-coupling integrals.  Changing the sign
  of $Q$ leads to two corresponding snake schemes. The symbols
  $I^+_0$, $I^-_0$, $I^+_1$, and $I^-_1$ indicate which part of the
  spin-orbit-coupling integrals are canceled
\label{fg:cancel8}}
\end{figure}

In the case of mirror symmetry in the ring
$I^+_{\frac{1}{4}}$ and $I^-_{\frac{1}{4}}$ are related by (\ref{eq:mir}),
\begin{eqnarray}
I^\pm_{\frac{1}{4}}&=&I^\pm_\frac{1}{8}
+(I^\pm_\frac{1}{8})^* \text{e}^{\text{i}(-\nu_0\pm Q)/4}\; ,\\
(I^\mp_{\frac{1}{4}})^*&=&(I^\mp_\frac{1}{8})^*
+I^\mp_\frac{1}{8} \text{e}^{-\text{i}(-\nu_0\mp Q)/4}
\nonumber\\
&=&I^\mp_{\frac{1}{4}} \text{e}^{-\text{i}(-\nu_0\mp Q)/4}\; .
\end{eqnarray}
With mirror symmetric quadrants the spin-orbit-coupling integral
then simplifies to
\begin{eqnarray}
I^\pm&=&I^\pm_{\frac{1}{4}}(1+\text{e}^{\text{i}(\alpha_1-\alpha_2\pm2Q/4)})
\nonumber\\
&+&
I^\mp_{\frac{1}{4}}(1+\text{e}^{\text{i}(-\alpha_2+\alpha_3\pm2Q/4)})
\text{e}^{\text{i}(\alpha_1\pm2Q/4)}
\end{eqnarray}
and again this additional symmetry does not simplify the compensation
of the spin-orbit integrals.
\paragraph{Schemes with 8 Snakes:}
The same procedure can now be repeated with 8 snakes.  For that
purpose 4 more horizontal Siberian Snakes are placed at the
locations $j \pi/2+\Delta\theta$, $j\in\{0,1,2,3\}$.  In terms
of the integrals 
\begin{eqnarray}
I^\pm_0
&=&
\int_{0}^{\Delta\theta}k_y
 \text{e}^{\text{i}(-\Psi\pm\Phi)}\text{d}\theta\; ,
\\
I^\pm_1
&=&
\int_{\Delta\theta}^{\frac{\pi}{2}}k_y
 \text{e}^{\text{i}(\Psi\pm\Phi)}\text{d}\theta\; ,
\end{eqnarray}
the spin orbit coupling integrals of (\ref{eq:socoupss}) are
\begin{eqnarray}
I^\pm &=& I^\pm_0(1+
 \text{e}^{\text{i}(-\Psi_0+\alpha_1+\Psi_1-\alpha_2
\pm(\Phi_{0}+\Phi_{1}))}+\ldots\\
&+&
 \text{e}^{\text{i}(-\Psi_0+\alpha_1+\Psi_1-\alpha_2\mp\ldots-\alpha_{n-2}
\pm(\Phi_{0}+\ldots+\Phi_{n-3}))})
\label{eq:8a}\nonumber\\
&+&
I^\pm_1 \text{e}^{\text{i}(-\Psi_0+\alpha_1\pm\Phi_{0})}
(1+
 \text{e}^{\text{i}(\Psi_1-\alpha_2-\Psi_2+\alpha_3\pm(\Phi_{1}+\Phi_{2}))}
\nonumber\\
&+&\ldots\nonumber\\
&+&
\text{e}^{\text{i}(\Psi_1-\alpha_2-\Psi_2+\alpha_3\pm\ldots+\alpha_{n-1}
\pm(\Phi_{1}+\ldots+\Phi_{n-2}))})\; .
\nonumber
\end{eqnarray}
If there is an additional mirror symmetry and the snakes are all
placed in the symmetry points, (\ref{eq:mir}) implies
$I^\pm_1=(I^\pm_0)^*\text{e}^{\text{i}(-\nu_0\pm Q)/4}$, which again
does not lead to simplifications. The complete spin phase advance of
the ring is $\sum_{j=0}^n(-)^j(\Psi_j+\alpha_j)=\pi$.  Since this
phase advance is required to be independent of energy,
$\sum_{j=0}^n(-)^j\Psi_j$ has to vanish.  Because of the
super-periodicity this requires $\Psi_0=\Psi_1$, and all the spin
phases $\Psi_j$ in the equations (\ref{eq:8a}) cancel.  Then in terms
of the difference angles $\Delta_{jk}=\alpha_j-\alpha_k$, spin
matching the ring requires
\begin{eqnarray}
1
&+&
 \text{e}^{\text{i}(\pm  Q/4+\Delta_{12})}+
 \text{e}^{\text{i}(\pm2 Q/4+\Delta_{12}+\Delta_{34})}
\nonumber\\
&+&
 \text{e}^{\text{i}(\pm3 Q/4+\Delta_{12}+\Delta_{34}+\Delta_{56})}=0\; ,
\label{eq:8c1}\\
1
&+&
 \text{e}^{\text{i}(\pm  Q/4-\Delta_{23})}+
 \text{e}^{\text{i}(\pm 2Q/4-\Delta_{23}-\Delta_{45})}
\nonumber\\
&+&
 \text{e}^{\text{i}(\pm 3Q/4-\Delta_{23}-\Delta_{45}-\Delta_{67})}=0\; .
\label{eq:8c2}
\end{eqnarray}
Sets of 4 complex numbers with modulus 1 can only add up to
zero by the three schemes shown in Fig.~\ref{fg:cancel}.  The
equations (\ref{eq:8c1}) and (\ref{eq:8c2}) have the same structure as
the matching conditions of equations (\ref{eq:4c}) and the relations
(\ref{eq:4ca}) can therefore be used to obtain the following two
ways to satisfy (\ref{eq:8c1}):
\begin{eqnarray}
\Delta_{12}\stackrel{\circ}=\Delta_{56}\stackrel{\circ}=\pi-Q/4\; &,&\ \
\Delta_{34}\stackrel{\circ}=3Q/4\; ,
\label{eq:a1}\\
\Delta_{12}\stackrel{\circ}=\Delta_{56}\stackrel{\circ}=\pi+Q/4\; &,&\
\ \Delta_{34}\stackrel{\circ}=-3Q/4\; .
\label{eq:a2}
\end{eqnarray}
Equation (\ref{eq:a2}) resulted from reversing the sign of $Q$ in
(\ref{eq:a1}).  There are also exactly two possibilities for
solving (\ref{eq:8c2}),
\begin{eqnarray}
\Delta_{23}\stackrel{\circ}=\Delta_{67}\stackrel{\circ}=\pi+Q/4\ &,&\
\ \Delta_{45}\stackrel{\circ}=-3Q/4\; ,
\label{eq:b1}\\
\Delta_{23}\stackrel{\circ}=\Delta_{67}\stackrel{\circ}=\pi-Q/4\ &,&\
\ \Delta_{45}\stackrel{\circ}= 3Q/4\; .
\label{eq:b2}
\end{eqnarray}
There are now 4 possibilities to snake match the ring; these are
obtained by combining the equations (\ref{eq:a1})\&(\ref{eq:b1}),
(\ref{eq:a1})\&(\ref{eq:b2}), (\ref{eq:a2})\&(\ref{eq:b1}), or
(\ref{eq:a2})\&(\ref{eq:b2}), where the last two possibilities
also result
from the first two by reversing the sign of $Q$. Figure
\ref{fg:cancel8} shows how parts of the ring cancel the depolarizing
effects of other parts in these snake matching schemes.

\begin{figure}[ht!]
\begin{center}
\begin{minipage}[c][0.5\columnwidth]{\columnwidth}
\includegraphics[width=0.49\columnwidth,bb=125 591 335 738]
                {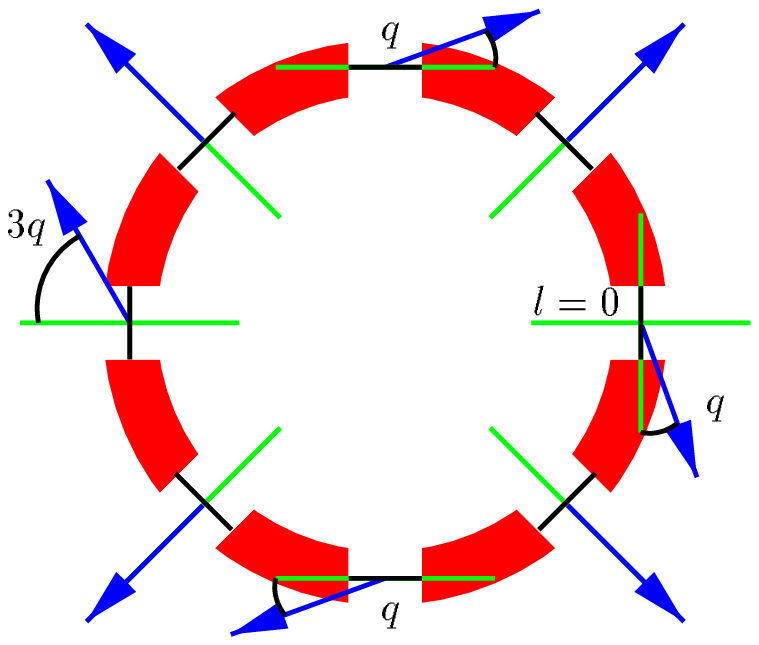}
\hfill
\includegraphics[width=0.49\columnwidth,bb=142 592 335 730]
                {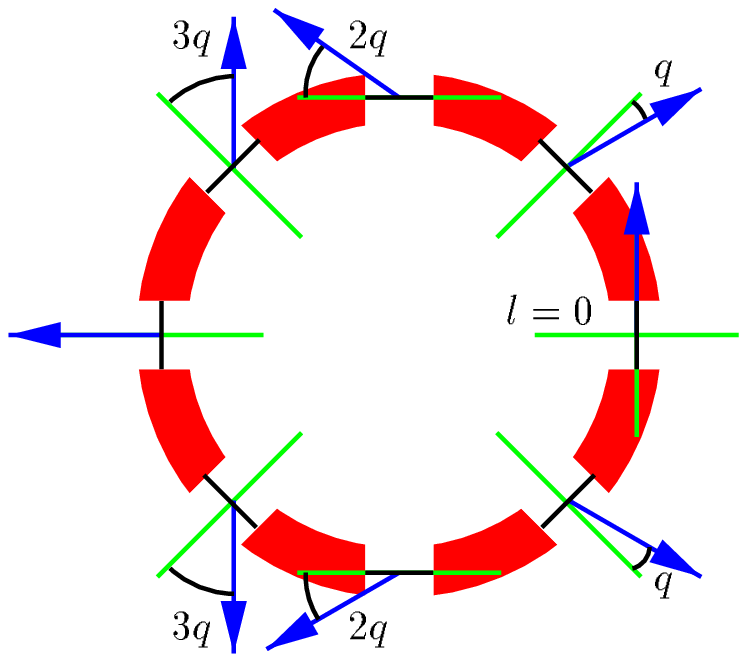}
\end{minipage}
\end{center}
\caption[Two energy independent snake matches with 8 Siberian Snakes]
 {Two of the 4
  ways to snake match a ring with super-periodicity 4 using 8
  horizontal Siberian Snakes.
  The number $q$ denotes $Q/8$ mod $\pi$ and lies
  in $[0,\pi)$.  The vertical tune times
  $2\pi$ is denoted by $Q$ and $\nu=G\gamma 2\pi$. The other two
  possible snake schemes are obtained by reversing the sign of $Q$.
  When all snake angles are increased by the same amount, then the
  ring remains spin transparent.  Note that the snake angle is
  independent of $\nu$ and thus of energy
\label{fg:8scheme}}
\end{figure}

Since only differences in the snake angles appear, one of the angles
can be chosen arbitrarily. This then fixes all other snake angles.
Here $\alpha_1=0$ is chosen for simplicity.  This leads to the
following possibilities:

Combination of the equations (\ref{eq:a1}) and (\ref{eq:b1}):
\begin{eqnarray}
\alpha_1&\stackrel{\circ}=&0    \; , \ \ \alpha_2\stackrel{\circ}=\pi+Q/4\; ,\ \
\alpha_3\stackrel{\circ}=0\; ,\\
\alpha_4&\stackrel{\circ}=&-3Q/4\; ,\ \ \alpha_5\stackrel{\circ}=0\; , \ \
\alpha_6\stackrel{\circ}=\pi+Q/4\; ,\\
\alpha_7&\stackrel{\circ}=&0    \; ,\ \ \alpha_8\stackrel{\circ}=\pi+Q/4\; .
\end{eqnarray}
Combination of the equations (\ref{eq:a1}) and (\ref{eq:b2}):
\begin{eqnarray}
\alpha_1&\stackrel{\circ}=&0    \; , \ \ \alpha_2\stackrel{\circ}=\pi+Q/4\; ,\ \
\alpha_3\stackrel{\circ}=2Q/4\; ,\\
\alpha_4&\stackrel{\circ}=&-Q/4\; ,\ \ \alpha_5\stackrel{\circ}=-4Q/4\; , \ \
\alpha_6\stackrel{\circ}=\pi-3Q/4\; ,\\
\alpha_7&\stackrel{\circ}=&-2Q/4\; ,\ \ \alpha_8\stackrel{\circ}=\pi-Q/4\; .
\end{eqnarray}
The values for $\alpha_8$ were obtained from the requirement
$\alpha_8-\alpha_7+\alpha_6-\alpha_5+\alpha_4-\alpha_3+\alpha_2-\alpha_1
\stackrel{\circ}=\pi$. The last snake scheme can be simplified by
decreasing all snake angles by $2Q/4$, leading to
\begin{eqnarray}
\alpha_1&\stackrel{\circ}=&Q/4\; , \ \ \alpha_2\stackrel{\circ}=\pi+2Q/4\; ,\ \
\alpha_3\stackrel{\circ}=3Q/4\; ,\\
\alpha_4&\stackrel{\circ}=&0\; ,\ \ \alpha_5\stackrel{\circ}=-3Q/4\; , \ \
\alpha_6\stackrel{\circ}=\pi-2Q/4\; ,\\
\alpha_7&\stackrel{\circ}=&-Q/4\; ,\ \ \alpha_8\stackrel{\circ}=\pi\; .
\end{eqnarray}
These snake schemes are shown in Fig.~\ref{fg:8scheme} where
account has been taken of the fact that the actual angle between the
snake's rotation axis and the radial direction is $\alpha/2$.
Furthermore advantage has been taken of the fact that the angle
$\alpha/2$ only needs to be known modulo $\pi$.

Here it is very important to note that the snake angles are independent
of $\nu=G\gamma$ and therefore that a snake match has been achieved for all
energies.  With 4 Siberian Snakes such an energy independent snake
match is not possible in a four-fold symmetric ring.

One could repeat the same procedure for a layout with 6
horizontal Siberian Snakes or with combinations of, for example, 6
horizontal Siberian Snakes and two type~III snakes.  Due to their
three-fold symmetry, this could be of special interest for the RHIC
rings.

\section{Snake Matching HERA-p}
\label{sc:snmatchhera}

\subsection{Schemes with 4 snakes:}
When the spin-orbit-coupling integrals starting at an azimuth
$\theta_0$ are minimized, the opening angle of the invariant spin
field at $\theta_0$ for the approximation of linear spin-orbit motion
is also minimized. In fact, $P_{lim}=1$ in linear approximation if
$I^\pm=0$ since then the spin motion is decoupled from the orbit to
first order.

It has been seen from the example of a
ring with super-periodicity 4 that 8 Siberian Snakes can be used to
snake match the spin-orbit-coupling integrals to zero at one azimuth
of the ring for all energies.  This could be achieved since the snake
angles were used to adjust the spin phase advances in such a way that
perturbations in one part of the ring were compensated by identical
perturbations in one of the identical super-periods of the ring.

HERA-p has 4 identical arcs separated by 4 not identical straight
sections. Due to the lack of four-fold symmetry it is in general not
possible to find snake angles which completely compensate all
spin-orbit-coupling integrals.  However, the 4 identical arc sections
of HERA dominate the spin-orbit resonance strength of vertical motion.
Thus it would make sense to arrange that the perturbing effect of
these arcs cancel each other.  A first step in this direction is a
symmetrization of the quadrants by making the spin phase advance in
all of the straight sections identical. This can be done by the
insertion of two more flattening snakes.

\begin{figure}[ht!]
\begin{center}
\begin{minipage}[t]{0.9\columnwidth}
\includegraphics[width=\columnwidth,bb=110 472 421 769,clip]
                {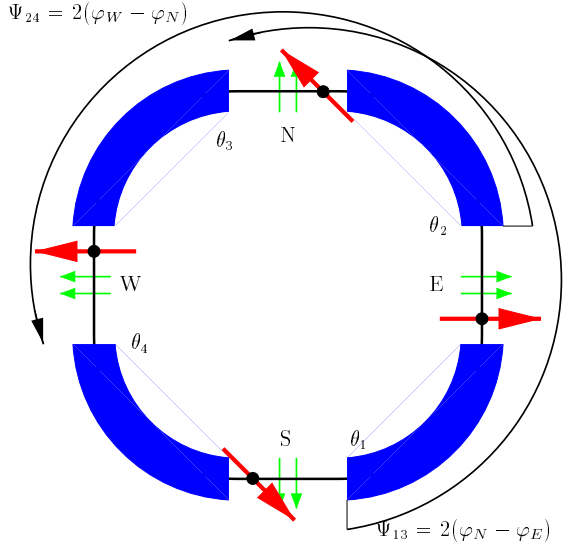}
\end{minipage}
\end{center}
\caption[Spin phase advances in a ring without exact
 super-periodicity] {The spin phase advance from the beginning of one
 regular arc to the beginning of the regular arc on the opposite side
 of the ring. Large arrows indicate Siberian Snakes, mall arrows
 indicate flattening snakes.
\label{fg:heramatch}}
\end{figure}

The spin-orbit-coupling integrals from the first regular FODO cell to
the last FODO cell of a regular arc in HERA-p will be denoted by
$\hat I_y^+$ and $\hat I_y^-$ and the azimuths of the beginnings of the
4 regular arcs as $\theta_1$, $\theta_2$, $\theta_3$, and $\theta_4$.
The central points of the South, West, North, and
East straight sections are
denoted by $S$, $W$, $N$, and $E$.  The spin phase advances
between the arcs are compensated using,
the snake angles $\varphi_E$, $\varphi_N$, and
$\varphi_W$.  The spin phase advance between $\theta_i$ and
$\theta_j$ is denoted by $\Psi_{ij}$.  These notations are indicated
in Fig.~\ref{fg:heramatch}.

With Siberian Snakes in each of the straight sections, the spin phase
advance from $\theta_1$ to $\theta_3$ is given by
$\Psi_{13}=\Psi_{1E}-2\varphi_E-\Psi_{EN}+2\varphi_N+ \Psi_{N3}$. The
spin phase advance is identical in all quadrants of the ring. The
total spin phase advance is then solely determined by the snake angles
and is therefore independent of energy:
$\Psi_{13}=2(\varphi_N-\varphi_E)$ and
$\Psi_{24}=2(\varphi_W-\varphi_N)$.  The orbital phase advance
$\Phi(\theta_3)-\Phi(\theta_1)$ also does not depend on energy.  For
simplicity, $\Phi(\theta_j)-\Phi(\theta_i)$ will now be denoted by
$\Phi_{ij}$.

The spin-orbit-coupling integrals at the South interaction point
then contain the following contributions from the
4 regular arcs:
\begin{eqnarray}
&&I^+_\mathrm{arcs}
= \hat I^+\text{e}^{\text{i}(-\Psi_{S1}+\Phi_{S1})}
(1+\text{e}^{\text{i}[2(\varphi_E-\varphi_N)+\Phi_{13}]})\\
&&+
(\hat I^-)^*
\text{e}^{\text{i}(2\varphi_E-\Psi_{SE}+\Psi_{E2}+\Phi_{S2})}
(1+\text{e}^{\text{i}[2(\varphi_W-\varphi_N)+\Phi_{24}]})
\nonumber\\
&&I^-_\mathrm{arcs} = \hat I^-\text{e}^{\text{i}(-\Psi_{S1}-\Phi_{S1})}
(1+\text{e}^{\text{i}[2(\varphi_E-\varphi_N)-\Phi_{13}]})\\
&&+
(\hat I^+)^*
\text{e}^{\text{i}(2\varphi_E-\Psi_{SE}+\Psi_{E2}-\Phi_{S2})}
(1+\text{e}^{\text{i}[2(\varphi_W-\varphi_N)-\Phi_{24}]})
\nonumber
\label{eq:ipmhera}
\end{eqnarray}

This shows that it is always possible to cancel one of the spin-orbit
coupling integrals by choosing the snake angles so that the spin
perturbation produced in one of the arcs is canceled by the arc on the
opposite side of the ring.  Since $|\hat I^+|$ and $|\hat I^-|$ are
different, neighboring arcs can in general not compensate each other
when 4 Siberian Snakes are used.

It is however possible to use the eight-snake scheme found for
symmetric lattices.  The two special 8 Siberian Snake schemes which
lead to an energy independent snake match in a ring with
super-periodicity 4 will not spin-match HERA-p completely, but the
spin perturbation from the arcs, which are the dominant perturbation,
will be compensated exactly.  This possibility of having a set of
Siberian Snake angles which do not have to be changed with energy and
which lead to a tightly bundled invariant spin field is on the one
hand very attractive; on the other hand it requires 8 Siberian Snakes
of which 4 would have to be installed at the centers of the HERA-p
arcs, where technical requirements of moving cryogenic feed-throughs
and super-conducting magnets would be very costly.  If possible, a
four-snake scheme should therefore be found.

Whereas it was shown below (\ref{eq:no4sol}) that a
four-snake scheme cannot cancel both spin-orbit-coupling integrals in
a ring with super-periodicity, a corresponding cancelation of the
spin perturbation due to the arcs in HERA-p can nevertheless be
achieved since the orbital phase advances between
the arcs can be manipulated individually, while these four phase
advances are equal for a lattice with super-periodicity 4.

To cancel both spin-orbit integrals in (\ref{eq:ipmhera}), 4
phase factors have to be $-1$.  This requires
\begin{eqnarray}
2(\varphi_E-\varphi_N)+\Phi_{13}&\stackrel{\circ}=&\pi\; ,\\
2(\varphi_E-\varphi_N)-\Phi_{13}&\stackrel{\circ}=&\pi\; ,\\
2(\varphi_W-\varphi_N)+\Phi_{24}&\stackrel{\circ}=&\pi\; ,\\
2(\varphi_W-\varphi_N)-\Phi_{24}&\stackrel{\circ}=&\pi\; .
\end{eqnarray}
Subtraction of the first two equations
leads to the requirement that the betatron phase advance from
$\theta_1$ half way around the ring to $\theta_3$ is an odd or even
multiple of $\pi$.  The same is true for the phase advance from
$\theta_2$ to $\theta_4$.  Correspondingly, the spin phase advance
over these regions has to be an odd multiple of $\pi$ when the orbit
phase advance is an even multiple and vice versa.  With a rather
benign change of the vertical optics in HERA-p which does not change
the vertical tune, the contribution of
the regular arcs to both spin-orbit-coupling integrals can be canceled,
even in a four-snake scheme.

\begin{figure}[ht!]
\begin{center}
\begin{minipage}[t]{\columnwidth}
\includegraphics[width=\columnwidth,bb=124 524 548 756,clip]
                {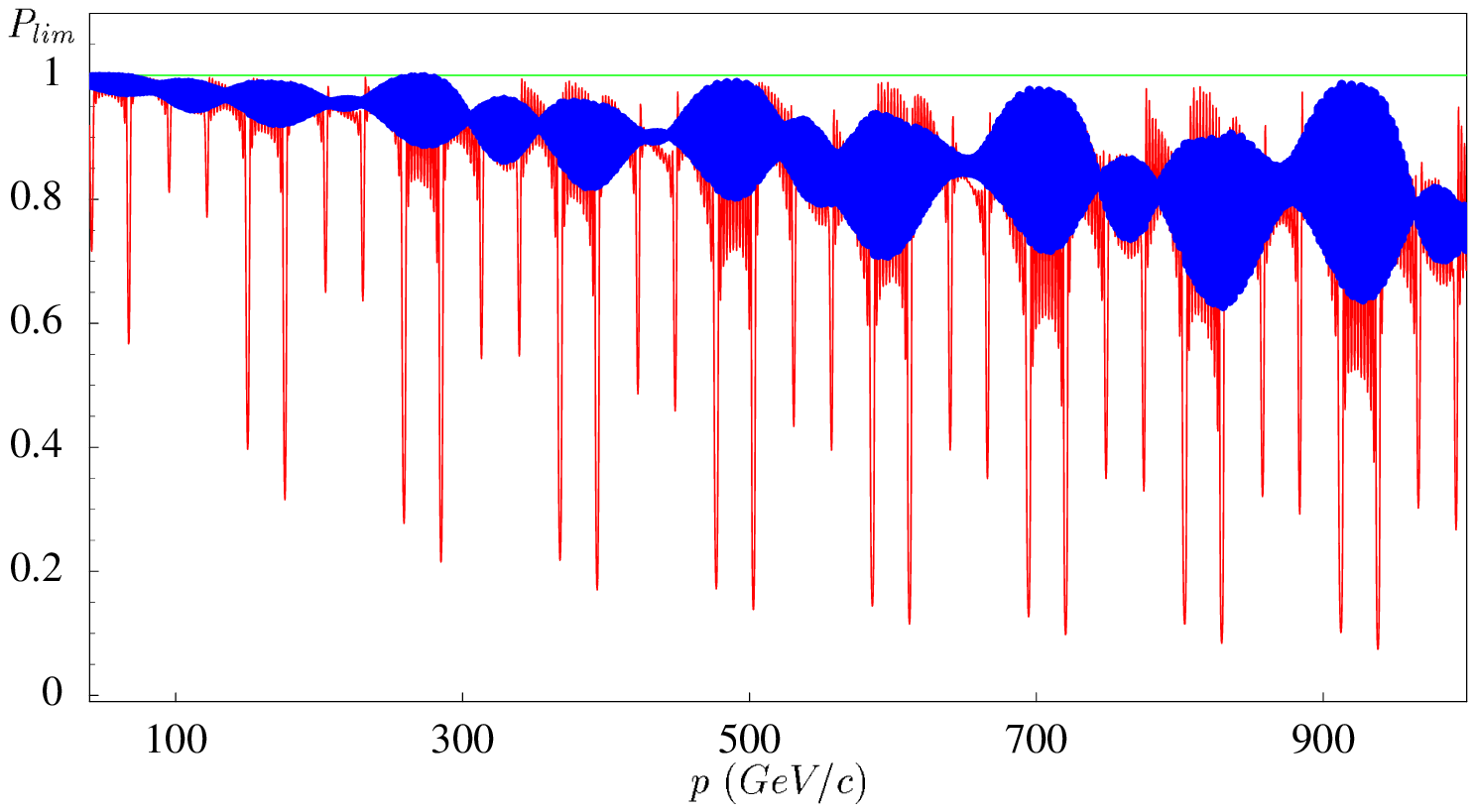}
\includegraphics[width=\columnwidth,bb=124 524 548 756,clip]
                {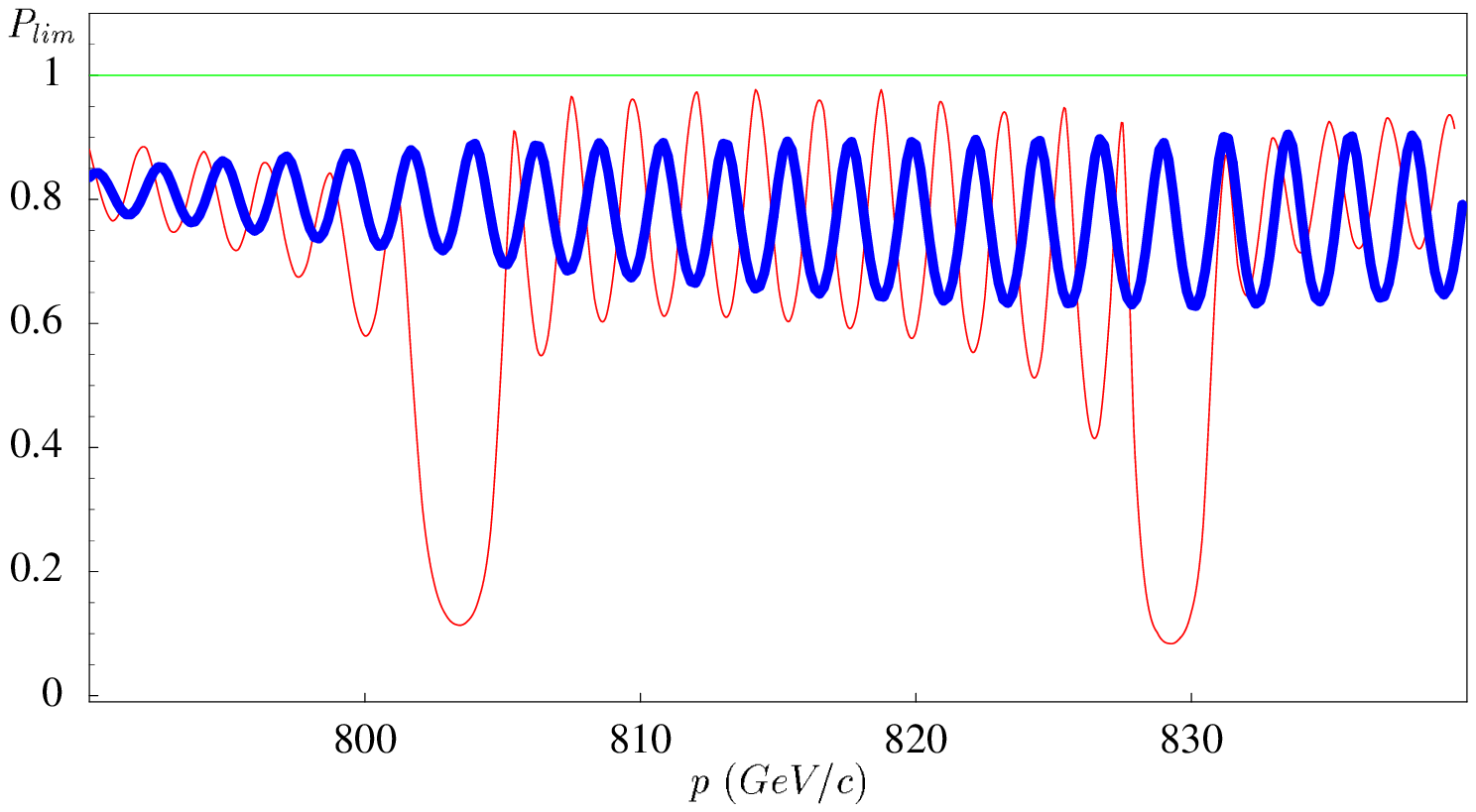}
\end{minipage}
\end{center}
\caption[Improvement of $P_{lim}$ by matching 4 snake angles
  and the orbital phases]
 {Improvement of $P_{lim}$ by matching 4 snake angles
  and the orbital phases.  The snake arrangement is
  $(0\frac{\pi}{2}\frac{\pi}{2}\frac{\pi}{2})o$ (\emph{blue}).  As a
  comparison $P_{lim}$ from linearized spin-orbit motion is shown for
  the same HERA-p optics with a
  $(\frac{\pi}{4}0\frac{\pi}{4}0)o$ snake scheme (\emph{red})
\label{fg:herasnmatch}}
\end{figure}

The snake scheme $(0\frac{\pi}{2}\frac{\pi}{2}\frac{\pi}{2})o$ has
$\Psi_{13}=0$ and $\Psi_{24}=0$.  For this snake scheme, the betatron
phase advances from $\theta_1$ to $\theta_3$ and from $\theta_2$ to
$\theta_4$ were adjusted to be odd multiples of $\pi$.  This change of
the optics is indicated by the index $o$ in the notation of the snake
scheme.  The maximum time average polarization $P_{lim}$ is plotted
(blue) in Fig.~\ref{fg:herasnmatch} for the complete range of HERA-p
momenta (top) and for the critical momentum regions above $800$~GeV/c
(bottom).  As a comparison, $P_{lim}$ for a standard snake scheme
$(\frac{\pi}{4}0\frac{\pi}{4}0)$ (red) is also shown.

The complete snake match of the arcs in HERA-p in deed
eliminates all strong reductions of $P_{lim}$ over the complete
momentum range.  Nonlinear effects will be analyzed later, but as far
as the linear effects are concerned, this snake matched lattice of
HERA-p would be a rather promising choice for the acceleration
of polarized proton beams.

\subsection{Schemes with 8 snakes:}
Although, for the reasons explained, eight snakes are not very
practical for HERA-p, significant improvements would be
possible if 8 snakes were used, as will now be shown.  The snake
matching schemes of Fig.~\ref{fg:8scheme} are suitable for rings with
super-periodicity 4.  They are therefore not directly applicable to
HERA-p.  However, the cancelation schemes of Fig.~\ref{fg:cancel8} can
still provide a guide to snake matching HERA-p with 8 snakes.

The 8 Siberian Snakes are placed in the straight sections and into the
centers of the arcs, so that the horizontal angle between adjacent
Siberian Snakes is $45^\circ$ and therefore $\Psi_j= G\gamma \pi/4$
for all $j\in\{0\ldots 7 \}$.  This ensures that the snake match is
independent of energy since then the $n$th octant's contribution
$I_{n-1}^\pm$ to the spin-orbit coupling integral $I^\pm$ can be
compensated by the 2nd neighbor's contributions $I_{n+1}^\pm$.  In
particular the $\Psi_j$ cancel in the phase factor in
\begin{eqnarray}
& &I_{n-1}^\pm\\
&+&I_{n+1}^\pm
\text{e}^{\text{i}[(-1)^n(\Psi_{n-1}-\alpha_n-\Psi_n+\alpha_{n+1})
\pm(\Phi_{n-1}+\Phi_{n})]}\ .\nonumber
\end{eqnarray}

Now a snake scheme for HERA-p is sought that is guided by the
cancelation scheme of Fig.~\ref{fg:cancel8}~(left).  When the first
octant's contribution $I_0^\pm$ to a spin-orbit coupling integral
$I^\pm$ is to be compensated by that of its 2nd neighbor $I_2^\pm$,
the phase factor in
\begin{equation}
I_0^\pm+\text{e}^{\text{i}[-\Psi_0+\alpha_1+\Psi_1-\alpha_2
             \pm(\Phi_{0}+\Phi_{1})]}I_2^\pm
\end{equation}
should be $-1$.  The same should be true for the phase factor in
\begin{equation}
I_4^\pm+\text{e}^{\text{i}[-\Psi_4+\alpha_5+\Psi_5-\alpha_6
             \pm(\Phi_{4}+\Phi_{5})]}I_6^\pm\ .
\end{equation}
These conditions are satisfied for the superscript $+$ as well as $-$
when all the snake angles $\alpha_j$ are zero for $j\in\{1,2,5,6\}$
and the betatron phase advances are chosen appropriately.  Similar
conditions arise for the phase factors involved in matching $I_1$
against $I_3$ and in matching $I_5$ against $I_7$.  They are also $-1$
when the betatron phases are chosen appropriately and when $\alpha_3$
and $\alpha_7$ are zero.  Since $\alpha_0$ and $\alpha_4$ do not
appear in these matching conditions, they can be chosen freely.  But
for a design orbit spin tune of $0.5$, one has to choose
$\alpha_0+\alpha_4=\pi$.  Here $\alpha_4=0$ has been chosen and this
snake scheme is then characterized as $(\frac{\pi}{2}0000000)o$.

In the case of HERA-p spin-orbit integrals of complete octants cannot
cancel, but the contribution of the regular arc sections can cancel.
Therefore the betatron phase advance is not chosen to be
$\Phi_0+\Phi_1\stackrel{\circ}=\pi$ between Siberian Snakes, but
rather the betatron phase advances between the beginning of the
regular arcs of the first and the second quadrant are chosen in that
way, i.e.~$\Phi_{12}\stackrel{\circ}=\pi$ with the notation of
Fig.~\ref{fg:heramatch}.  In this configuration the regular arc of the
first octant cancels that of the third, i.e.~the regular arc's
contribution to $I_2$ cancels that to $I_0$.  But also the regular arc
of the second octant cancels that of the fourth since the betatron
phase advance between the centers of the first and the second arc is
then also $\pi$ mod $2\pi$, i.e.~the regular arc part of $I_3$ cancels
that of $I_1$.  Similarly the octants of the third quadrant cancel
those of the fourth quadrant.  In the snake scheme
$(\frac{\pi}{2}0000000)o$ the phase advance over the East straight
section was changed by $2\pi\times 0.1028$ to have
$\Phi_{12}=2\pi\times 8.5$.  The phase advance over the West was
changed by $2\pi\times 0.0208$ to have $\Phi_{34}=2\pi\times 7.5$.  In
order to have the same betatron phase advance in the North and the
South straight sections and to keep the total tune constant, the
linear optics of the North and of the South straight sections were
modified.

This cancelation scheme does not agree completely with that of Fig.
\ref{fg:cancel8}~(left) since in $(\frac{\pi}{2}0000000)o$ the
superscripts $+$ and $-$ have been dealt with simultaneously so that
$I_0^\mp$ no longer has to compensate $I_4^\mp$.  Since
$\Phi_{12}\stackrel{\circ}=\pi$ and $\Phi_{34}\stackrel{\circ}=\pi$,
it is appropriate to take $q=\pi/2$ for the phase advance of one
octant of the scheme in Fig.~\ref{fg:8scheme}~(left), for which one
then finds a close similarity to the snake scheme
$(\frac{\pi}{2}0000000)o$.  The resemblance would be even closer if
the phase advance over the North straight section had been changed so
that $\Phi_{23}\stackrel{\circ}=0$ and if $\alpha_4=\pi$, $\alpha_0=0$
had been chosen.  This would require an additional change of
$\Phi_{41}$ to adjust the orbital tune.

A snake scheme for HERA-p which resembles
Fig.~\ref{fg:cancel8}~(right) can also be found.  In this cancelation
scheme the phase factors in the following sums have to be $-1$:
\begin{eqnarray}
I_0^+ &+& \text{e}^{\text{i}[-\Psi_0+\alpha_1+\Psi_1-\alpha_2
+(\Phi_{0}+\Phi_{1})]}I_2^\pm \\
I_4^+ &+&
\text{e}^{\text{i}[-\Psi_4+\alpha_5+\Psi_5-\alpha_6
+(\Phi_{4}+\Phi_{5})]}I_6^\pm \\
I_0^- &+&
\text{e}^{\text{i}(-\Psi_0+\alpha_1+\Psi_1-\alpha_2 -
\Psi_2+\alpha_3+\Psi_3-\alpha_4)}
\nonumber\\
&\times&
\text{e}^{-\text{i}(\Phi_{0}+\Phi_{1}+\Phi_{2}+\Phi_{3})}I_4^\pm \\
I_2^- &+&
\text{e}^{\text{i}(-\Psi_2+\alpha_3+\Psi_3-\alpha_4-\Psi_4 +
\alpha_5+\Psi_5-\alpha_6)}
\nonumber\\
&\times&
\text{e}^{-\text{i}(\Phi_{2}+\Phi_{3}+\Phi_{4}+\Phi_{5})}I_6^\pm \\
I_1^- &+&
\text{e}^{\text{i}[ \Psi_1-\alpha_2-\Psi_2+\alpha_3
-(\Phi_{1}+\Phi_{2})]}I_3^\pm \\ I_5^- &+& \text{e}^{\text{i}[
\Psi_5-\alpha_6-\Psi_6+\alpha_7 -(\Phi_{5}+\Phi_{6})]}I_7^\pm \\
I_1^+ &+& \text{e}^{\text{i}(
\Psi_1-\alpha_2-\Psi_2+\alpha_3+\Psi_3-\alpha_4-\Psi_4+\alpha_5)}
\nonumber\\
&\times&
\text{e}^{\text{i}(\Phi_{1}+\Phi_{2}+\Phi_{3}+\Phi_{4})}I_5^\pm \\
I_3^+ &+&
\text{e}^{\text{i}(
\Psi_3-\alpha_4-\Psi_4+\alpha_5+\Psi_5-\alpha_6-\Psi_6+\alpha_7)}\nonumber\\
&\times&
\text{e}^{\text{i}(\Phi_{3}+\Phi_{4}+\Phi_{5}+\Phi_{6})}I_7^\pm
\end{eqnarray}

\begin{figure}[ht!]
\begin{center}
\begin{minipage}[t]{\columnwidth}
\includegraphics[width=\columnwidth,bb=124 524 548 763,clip]
                {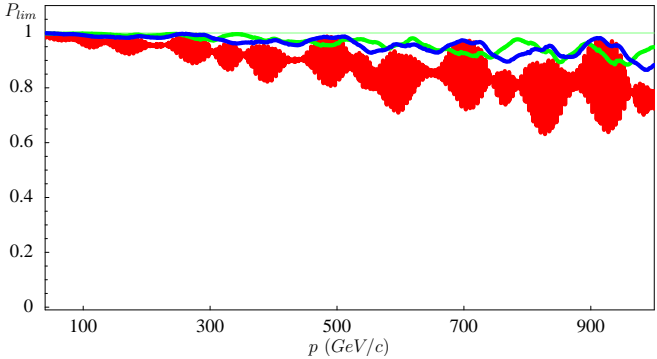}
\end{minipage}
\end{center}
\caption[Improvement of $P_{lim}$ by matching 8 snake angles
and the orbital phases] {Improvement of $P_{lim}$ by matching 8 snake
  angles and the orbital phases.  The snake arrangement is
  $(\frac{\pi}{2}0000000)o$ (\emph{blue}) and
  \mbox{$(\frac{\pi}{2}abc0$-$c$-$b$-$a)o$} (\emph{green}).  As a
  comparison $P_{lim}$ from linearized spin-orbit motion is shown
  for the snake and phase advance matched four-snake scheme
  $(0\frac{\pi}{2}\frac{\pi}{2}\frac{\pi}{2})o$ (\emph{red})
\label{fg:plim8}}
\end{figure}

Again complete octants cannot cancel, but to cancel the contributions
of regular arc sections, the beginning of a regular arc is chosen as
the starting point for the compensation.  Now the $\Phi_{n}$ for all
even $n$ describe the betatron phase advance from the beginning to the
center of the regular arc, and they are therefore equal.  For the
choice $\alpha_4=0$ all phase factors are $-1$ if and only if
$\Phi_{5} = \Phi_{1}$ and
\begin{eqnarray}
&& \alpha_1 \stackrel{\circ}=  \Phi_{23}\; ,\ \
   \alpha_2 \stackrel{\circ}=  \Phi_{13}\; ,\ \
   \alpha_3 \stackrel{\circ}=  \Phi_{14}\; ,\\
&& \alpha_5 = -\alpha_3\; ,\ \
   \alpha_6 = -\alpha_2\; ,\ \
   \alpha_7 = -\alpha_1\; .
\end{eqnarray}
This snake scheme is referred to as
\mbox{$(\frac{\pi}{2}abc0$-$c$-$b$-$a)o$}.  The condition
$\Phi_{5}=\Phi_{1}$ was satisfied by changing the betatron phase
advance over the East and West straight section without changing the
vertical tune.

Linearized spin orbit motion leads to a very favorable $P_{lim}$ for
both eight-snake schemes as shown in Fig.~\ref{fg:plim8}, where it is
compared to the $P_{lim}$ of the snake and phase advance matched
four-snake scheme $(0\frac{\pi}{2}\frac{\pi}{2}\frac{\pi}{2})o$.

\section{Higher-Order Resonances and Snake Schemes}
\label{sc:sshores}

At the critical energies, where the maximum time average polarization
is low during the acceleration process, linearized spin-orbit motion
does not describe spin dynamics well.  The spin motion is influenced
by several overlapping resonances in these regions and the single
resonance approximations \cite{froissart60,hoffstaetter99i} can also
not be applied.  Thus the simulation results obtained with these
computationally quick techniques should always be checked with more
time consuming non-perturbative methods if possible.  This is also
true for the snake-matched lattices of HERA-p with large $P_{lim}$,
even though they avoids large variations of the invariant spin field
$\vec n(\vec z)$ over the phase space of the beam in linearized
spin-orbit motion.  When first-order effects are canceled, the
higher-order effects become dominant and the quality of the
snake-matched lattice of HERA-p can only be evaluated with
higher-order theories.

Until 1996, when stroboscopic averaging \cite{hoffstaetter96d} was
introduced, there was no non-perturbative method of computing the
$\vec n$-axis at high energy in proton storage rings, where
perturbative methods are usually not sufficient
\cite{hoffstaetter99k}.  In addition, the method of anti-damping was
derived \cite{hoffstaetter98d}, which also computes $\vec n(\vec z)$
non-perturbatively and which can be faster when the $\vec n$-axis is
required for a range of phase space amplitudes.  Both methods of
computing the invariant spin field are implemented in the spin-orbit
dynamics code SPRINT, by which also the amplitude-dependent spin tune
$\nu(\vec J)$ can be computed once $\vec n(\vec z)$ is known.  Since
stroboscopic averaging and anti-damping are based on multi-turn
tracking data, they are applicable to all kinds of circular
accelerators and they are especially efficient for small rings and for
simple model accelerators. Nonlinear motion for orbit and spin
coordinates can esily be included~\cite{hoffstaetter00a, hoffstaetter99h}.

Subsequently another non-perturbative algorithm for computing $\vec
n(\vec z)$ and $\nu(\vec J)$ has been derived \cite{yokoya99}.  It is
called SODOM-2 since it was inspired by the earlier algorithm SODOM
\cite{yokoya92} which for convergence required the angle between $\vec
n$ and $\vec n_0$ to be small.  With some routines provided by
K.~Yokoya, SODOM-2 was incorporated into the program SPRINT
\cite{man_sprint02,hoffstaetter96d,hoffstaetter96h} and leads to
results which agree very well with those of stroboscopic averaging.
For motion in one degree of freedom, SODOM-2 is often faster than
stroboscopic averaging, especially for large rings like HERA-p where
particle tracking is relatively time consuming.  But for orbit motion
in more than one degree of freedom or in the vicinity of spin-orbit
resonances, SODOM-2 becomes exceedingly slow and then stroboscopic
averaging and anti-damping are needed.

\begin{figure}[ht!]\begin{center}
\begin{minipage}[t]{\columnwidth}
\includegraphics[width=0.49\columnwidth,bb=126 643 353 771,clip]
             {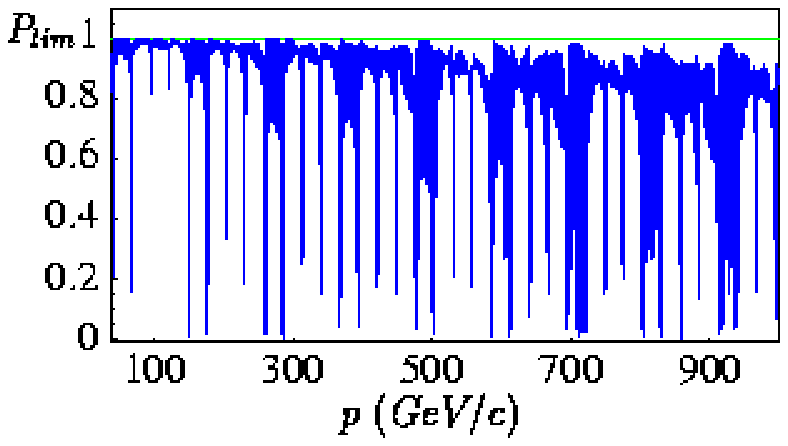}
\includegraphics[width=0.49\columnwidth,bb=113 630 336 755,clip]
             {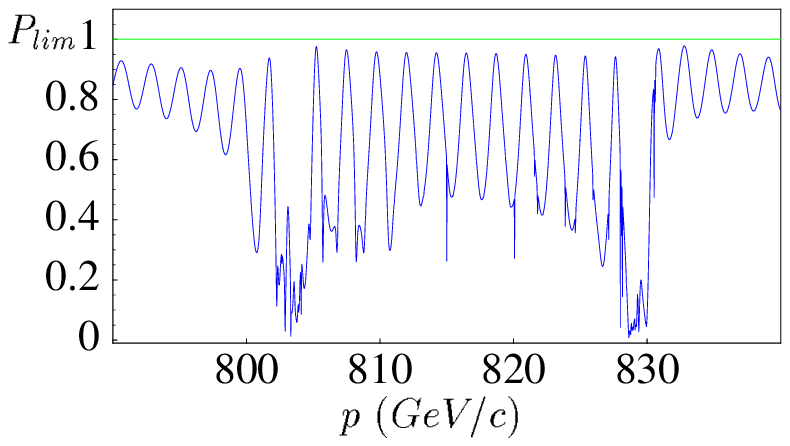}
\includegraphics[width=0.49\columnwidth,bb=126 643 351 771,clip]
             {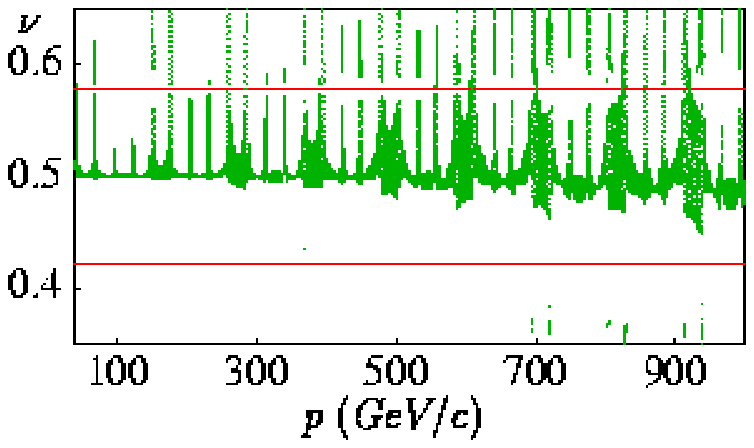}
\includegraphics[width=0.49\columnwidth,bb=113 630 336 755,clip]
             {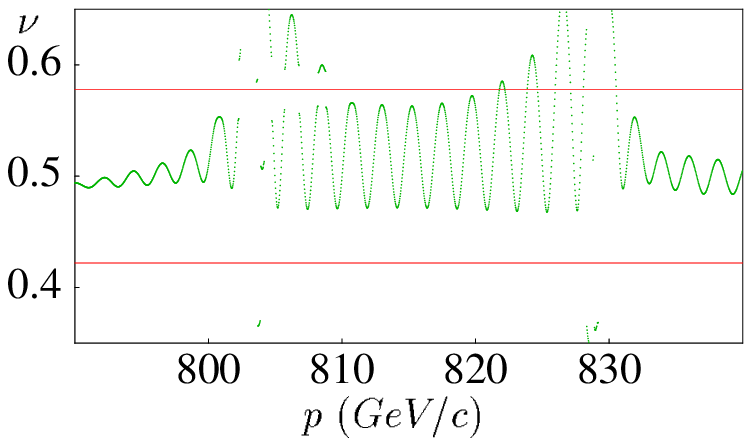}
\end{minipage}
\end{center}
\caption[$P_{lim}$ and $\nu$ with higher-order effects for a standard
         snake scheme]
 {$P_{lim}$ and $\nu$ for particles with a vertical amplitude
  corresponding to the $2.5\sigma$ emittance in the HERA-p
  lattice of the year 2002 with the
  $(\frac{\pi}{4}0\frac{\pi}{4}0)$ scheme.  {\bf Top}: The
  maximum time average polarization $P_{lim}$ for the complete
  acceleration range (\emph{left}) and for the critical energy range above
  800~GeV/c which has to be crossed when accelerating to the proposed
  storage energy of 870~GeV/c.  {\bf Bottom}: the corresponding amplitude
  dependent spin tune $\nu(J_y)$.  The second-order resonances
  $\nu=2Q_y$ and $\nu=1-2Q_y$ are indicated (\emph{red})
\label{fg:1b1b6fs}}
\end{figure}

To check whether the improvements of spin motion obtained in the
framework of linearized spin-orbit motion survive when higher-order
effects are considered, $P_{lim}$ and $\nu$ has been calculated.  The
result for one of the standard Siberian Snake schemes which used to be
considered advantageous by popular opinion is shown for the South
interaction point of HERA-p in Fig.~\ref{fg:1b1b6fs}.  It has four
Siberian Snakes in the $(\frac{\pi}{4}0\frac{\pi}{4}0)$ scheme.  Some
of the features of $P_{lim}$ were already revealed by linearized
spin-orbit motion in Fig.~\ref{fg:slimherast}.  Now many higher-order
resonances are revealed, causing strong reduction of $P_{lim}$ and
there are corresponding strong variations of the amplitude-dependent
spin tune $\nu$ \cite{spin00,epic00}.  Strong resonances occur
especially in the critical energy region where linearized spin orbit
motion in Sect.~\ref{sc:optsnake} already indicated a very small
$P_{lim}$ due to a coherent spin perturbation in all regular FODO
cells.  Many higher-order resonances overlap in these critical energy
regions of Fig.~\ref{fg:1b1b6fs}~(top) where large spin tune jumps can
be observed in Fig.~\ref{fg:1b1b6fs}~(bottom).  The strongest spin
tune jumps occur in the critical energy regions, mostly at the second
order resonance $\nu=2Q_y$ which is indicated by the top line
\cite{hoffstaetter99f}.

\begin{figure}[ht!]
\begin{center}
\begin{minipage}[t]{\columnwidth}
\includegraphics[width=0.49\columnwidth,bb=113 630 336 755,clip]
             {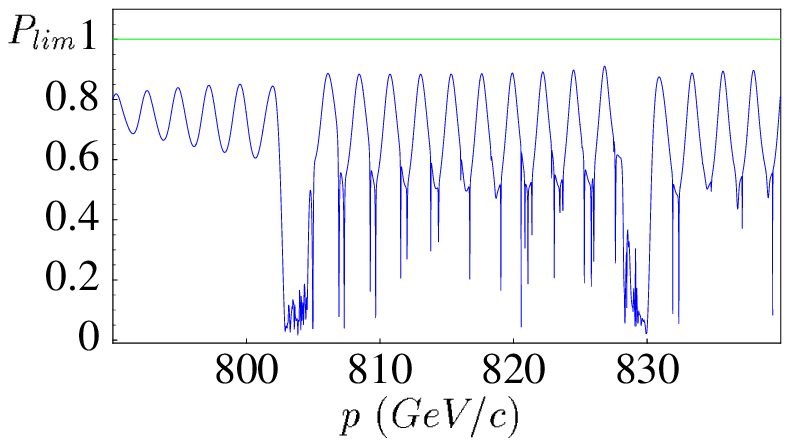}
\includegraphics[width=0.49\columnwidth,bb=113 630 336 755,clip]
             {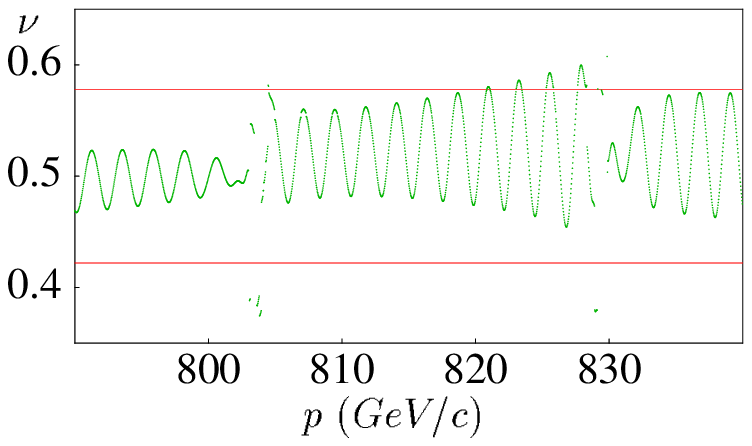}
\includegraphics[width=\columnwidth,bb=138 536 565 773,clip]
             {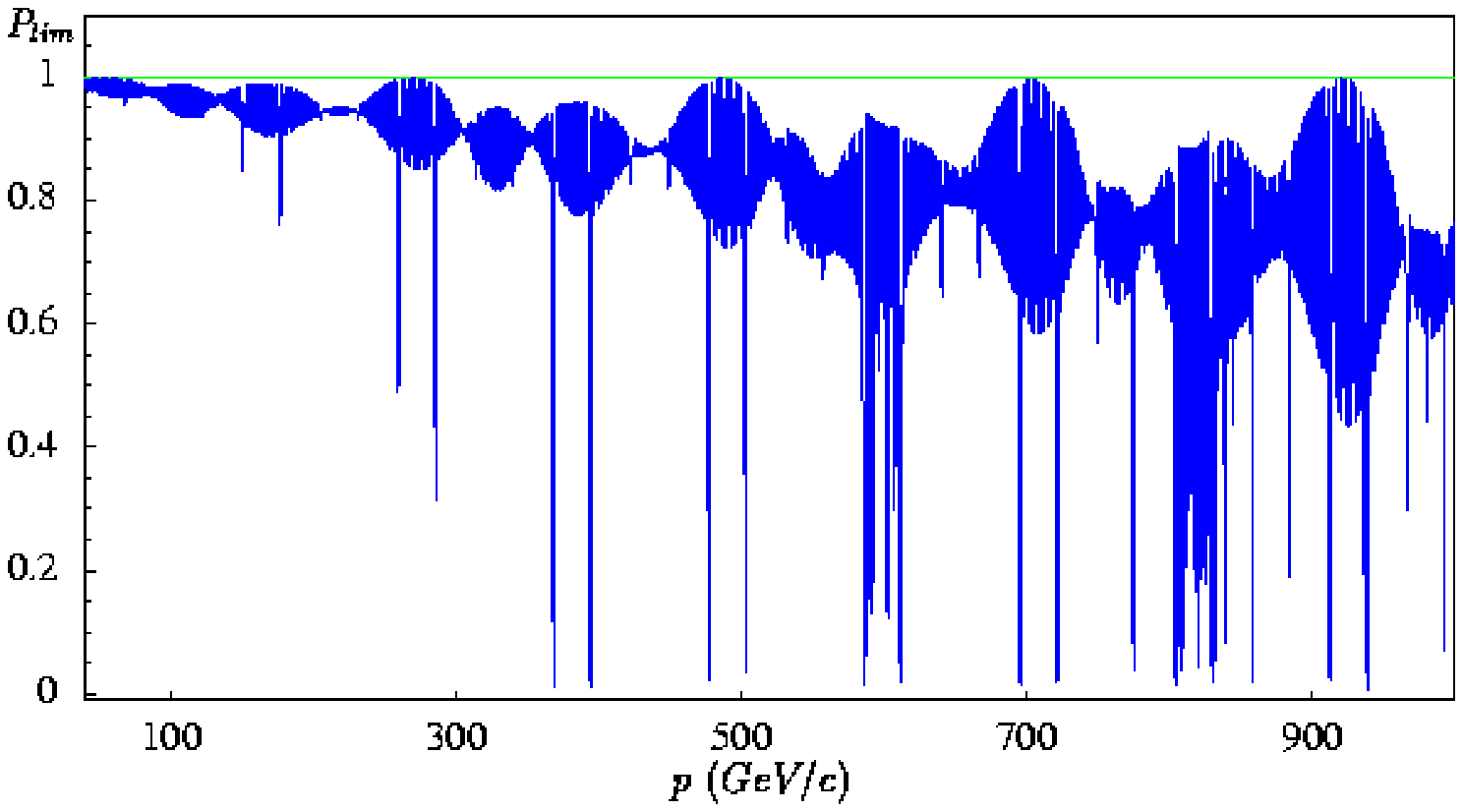}
\includegraphics[width=\columnwidth,bb=138 536 565 773,clip]
             {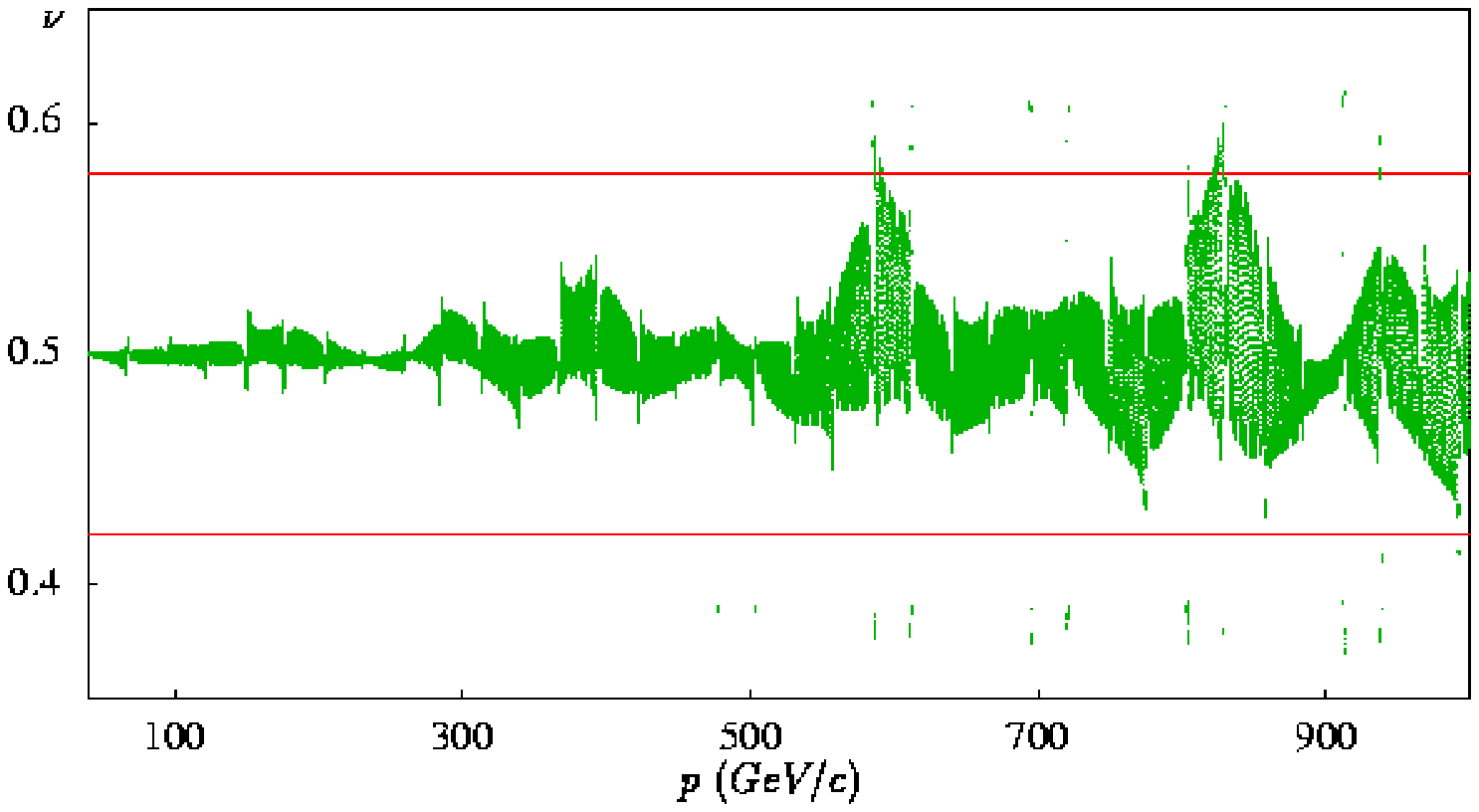}
\end{minipage}
\end{center}
\caption[$P_{lim}$ and $\nu$ with higher-order effects for the
  snake matched scheme of 4 Siberian Snakes]
 {$P_{lim}$ and $\nu$ for a $2.5\sigma$ vertical amplitude after the betatron
  phase advance between opposite regular arc structures was adjusted
  to be an odd multiple of $\pi$ in the
  $(0\frac{\pi}{2}\frac{\pi}{2}\frac{\pi}{2})o$ scheme. The
  resonances $\nu=2Q_y$ and $\nu=1-2Q_y$ are indicated (\emph{red})
\label{fg:21224as}}
\end{figure}

Figure \ref{fg:21224as} also shows $P_{lim}$ and $\nu$ for
higher-order spin dynamics in the snake-matched and
phase-advance-matched HERA-p ring with 4 Siberian Snakes.  While the
overall behavior of $P_{lim}$ over the complete acceleration range of
HERA-p looks similar to the result obtained with linearized spin-orbit
motion, which was displayed in Fig.~\ref{fg:herasnmatch}, higher-order
effects become very strong at high energies, especially in the
vicinity of the critical energies where perturbations of spin motion
in each FODO cell accumulate.  The spin tune spread at momenta below
400~GeV/c is small and higher-order effects seem to be benign even at
these critical energies.  A comparison of Figs.~\ref{fg:1b1b6fs} and
\ref{fg:21224as} hello shows that the special scheme obtained by
matching orbital phases and snake angles would be a very good choice
for HERA-p up to 300 or 400~GeV/c.  For linearized spin-orbit motion,
this snake-matched scheme does not produce a strong reduction of
$P_{lim}$ at any critical energy, but the higher-order effects become
very pronounced at some top energies of HERA-p.  Nevertheless, the
advantage over other snake schemes becomes clear in figure
Fig.~\ref{fg:21224as}~(bottom) where the amplitude dependent spin tune
$\nu$ is shown.  It comes close to a second order resonance at fewer
places and does not exhibit spin tune jumps which are as strong as
those in previous figures of $\nu$.

In this snake-matched scheme, the influence of higher-order effects
can be seen very clearly, because the first-order effects have been
matched to be very small.  The analysis of $\nu$ shows that completely
snake matching the spin perturbations in the arcs of HERA-p with 4
Siberian Snakes is advantageous, even though dips of $P_{lim}$ due to
higher-order resonances can be observed at high energies.  Even
around 300~GeV/c there are resonant dips of $P_{lim}$ in
Fig.~\ref{fg:21224as}~(middle) but they are less pronounced than those
in Fig.~\ref{fg:1b1b6fs} so that the snake matched scheme should be
very advantageous in the complete energy range of HERA-p.

\section{Polarization Reduction During Acceleration}

It should be noted that the destructive spin tune jumps at second
order resonances disappear completely when HERA-p is simulated without
its non-flat regions.  This is due to the fact that a large class of
resonances are not excited at all in mid-plane symmetric rings
\cite{vogt00,hoff04}. To reduce these perturbations, the East region
of HERA-p will now be simulated as flat since the HERMES experiment
located in this region does not require that the proton beam is on the
level of the electron beam.

\begin{figure}[ht!]
\begin{center}
\begin{minipage}[t]{\columnwidth}
\includegraphics[width=0.32\columnwidth,bb=113 599 266 756,clip]
             {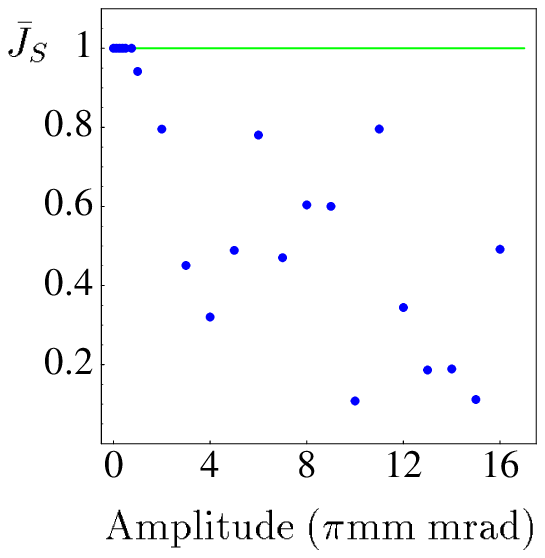}
\includegraphics[width=0.32\columnwidth,bb=113 599 266 756,clip]
             {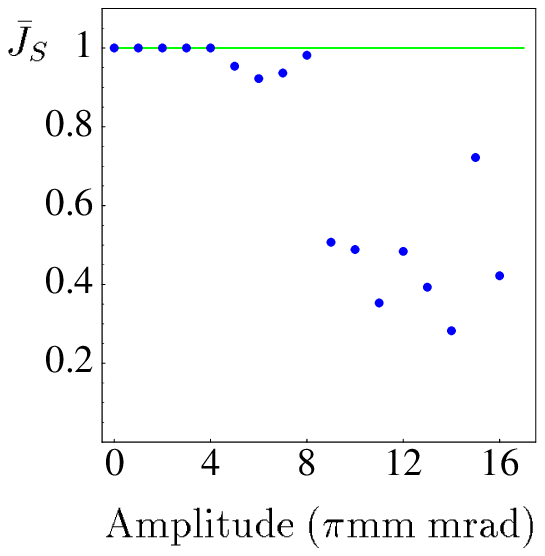}
\includegraphics[width=0.32\columnwidth,bb=113 599 266 756,clip]
             {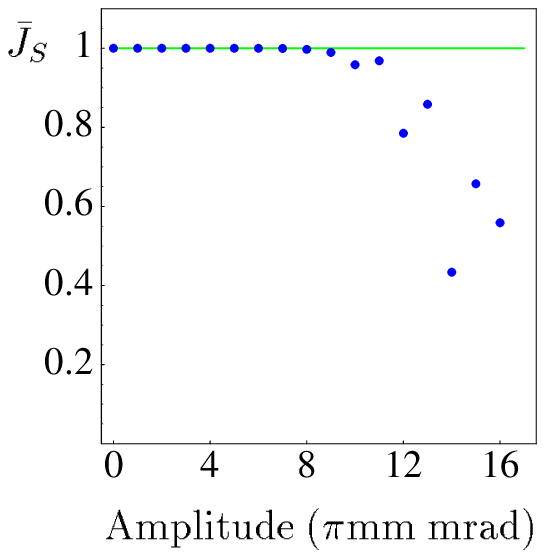}
\end{minipage}
\end{center}
\caption[Polarization reduction during acceleration for 3
         four-snake schemes] 
 {Average spin action $\bar J_{S}$ at 806~GeV for
  particles starting with $J_{S}=1$ at 800~GeV/c for three
  different snake schemes. {\bf Left}: $(\frac{\pi}{4}0\frac{\pi}{4}0)$,
  {\bf Middle}: $(\frac{3\pi}{4}\frac{3\pi}{8}\frac{3\pi}{8}\frac{\pi}{4})o$,
  {\bf Right}:  $(0\frac{\pi}{2}\frac{\pi}{2}\frac{\pi}{2})o$ scheme.
  Particles with an amplitude above 1 (left), 4 (middle), and 8 (right)
  lead to a reduction of polarization when the beam is accelerated through
  this critical energy region
\label{fg:4sramp}}
\end{figure}

When a particle is accelerated across the critical momentum region
from 800 to 806~GeV/c with a typical acceleration rate of 50~keV per
turn, the adiabatic invariance of $J_S=\vec n\cdot\vec S$ can be
violated and the level of violation will depend on the orbital
amplitude and the snake scheme.  This violation is illustrated in the
graphs in Fig.~\ref{fg:4sramp} which, for three different snake
schemes, show the average spin action $\bar J_{S}$ at 806~GeV/c which
had initially $J_{S}=1$ at 800~GeV/c before acceleration.

The change of $J_S$ in the critical energy region depends on the
initial phase space angle so that if $J_S$ had been computed only for
one particle, it could by chance have had an angle variable for which
$J_S$ does not change although it would have changed for other points
with the same vertical phase space amplitude.  To avoid such a chance
effect which gives the impression that $J_S$ is invariant, three
particles were accelerated and the average $\bar J_S$ is displayed in
Fig.~\ref{fg:4sramp}.

At small phase space amplitudes, $J_S$ is nearly invariant and
therefore $\bar J_S=1$.  For each of the three snake schemes, there is
a phase space amplitude $J_{ymax}$ above which $\bar J_S<1$ and the
regions of the beam with an amplitude above $J_{ymax}$ lead to a
reduction of the beam's polarization during the acceleration process.

For the standard snake scheme $(\frac{\pi}{4}0\frac{\pi}{4}0)$,
only the part of the beam with less than 1$\pi$~mm~mrad vertical
amplitude can remain polarized.  For the filtered scheme
$(\frac{3\pi}{4}\frac{3\pi}{8}\frac{3\pi}{8}\frac{\pi}{4})o$, phase
space amplitudes up to 4$\pi$~mm~mrad are allowed.  Finally the snake
matched scheme $(\frac{\pi}{4}0\frac{\pi}{4}\frac{\pi}{4})o$ gives
the most stable spin motion and Fig.~\ref{fg:4sramp} shows that
vertical amplitudes of up to 8$\pi$~mm~mrad are allowed.
This shows that the snake matched scheme is superior to the other
four-snake schemes studied here.  It stabilizes spin motion for 10
times larger phase space amplitudes than some other snake schemes.
Nevertheless, 8$\pi$~mm~mrad is not enough to allow high polarization
at top energies for today's emittances in HERA-p.

\begin{figure}[ht!]
\begin{center}
\begin{minipage}[t]{\columnwidth}
\includegraphics[width=0.49\columnwidth,bb=115 628 336 755,clip]
  {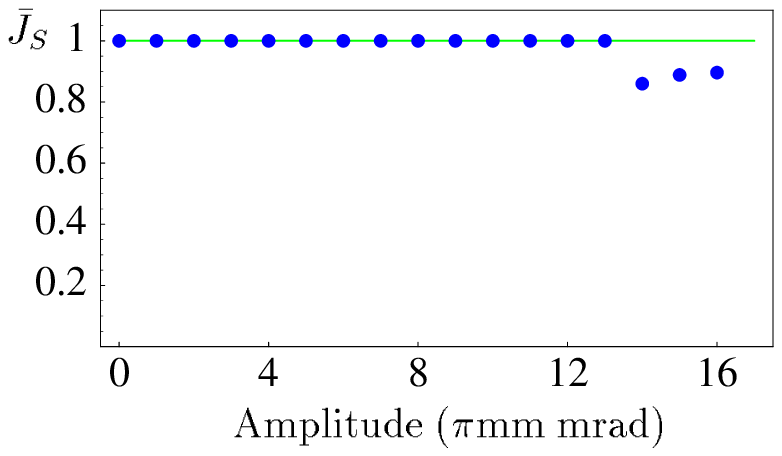}
\includegraphics[width=0.49\columnwidth,bb=115 628 336 755,clip]
  {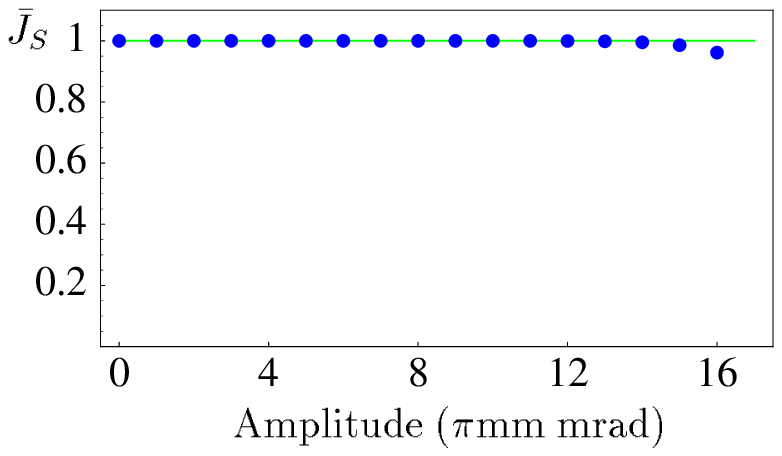}
\end{minipage}
\end{center}
\caption[Polarization reduction during acceleration for 2
eight-snake schemes] {The average spin action $\bar J_{S}$ at
  806~GeV for particles which started with $J_{S}=1$ at 800~GeV/c for
  the 2 different snake matched eight-snake schemes
  $(\frac{\pi}{2}0000000)o$ (left) and
  \mbox{$(\frac{\pi}{2}abc1$-$c$-$b$-$a)o$} (right).  Particles with
  an amplitude above 13 (left) and 15 (right) lead to a reduction of
  polarization when the beam is accelerated through this critical
  energy region
\label{fg:8sramp}}
\end{figure}

Matching 8 Siberian Snakes in HERA-p lead to two snake schemes with
very high $P_{lim}$ and very small spin tune spread.  But to
demonstrate that it is possible to further stabilize spin motion in
HERA-p by such schemes, Fig.~\ref{fg:8sramp} shows the vertical phase
space amplitudes for which $J_S$ remains invariant.  The more
effective of the two snake scheme stabilizes spin motion up to a
vertical amplitude of $14\pi$~mm~mrad.

\begin{figure}[ht!]
\begin{center}
\begin{minipage}[t]{\columnwidth}
\includegraphics[width=\columnwidth,bb=124 517 549 755,clip]
  {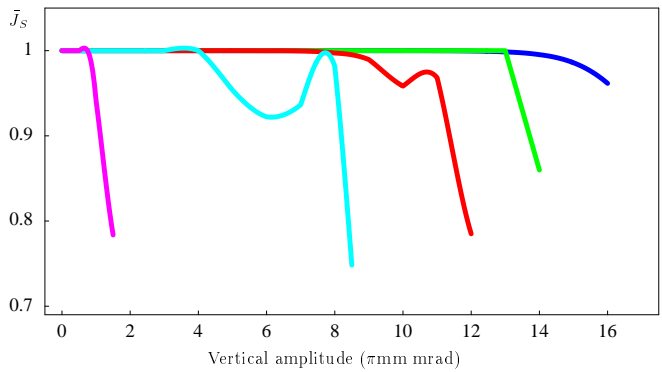}
\end{minipage}
\end{center}
\caption[Polarization reduction during acceleration for schemes with 4
and 8 Siberian Snakes] {The average spin action $\bar J_S$ at 806~GeV
for particles which started initially with $J_S=1$ at 800~GeV/c.
Violet: The standard scheme which stabilizes spin motion for particles
within 1$\pi$~mm~mrad, Cyan: the filtered four-snake scheme stabilizes
within 4$\pi$~mm~mrad, Red: the snake-matched four-snake scheme
stabilizes within 8$\pi$~mm~mrad, Green: the snake-matched eight-snake
scheme which stabilizes within 13$\pi$~mm~mrad, Blue: and the
snake-matched eight-snake scheme which stabilizes within
14$\pi$~mm~mrad of vertical phase space amplitude
\label{fg:allsramp}}
\end{figure}

These results for the various snake schemes are collected in
Fig.~\ref{fg:allsramp}, where it becomes clear that snake matching
with 4 and especially with 8 snakes leads to a significant
improvement.  In HERA-p it does not suffice to avoid a reduction of
$J_S$ for particles with less than $14\pi$~mm~mrad amplitude. It would
therefore be very helpful to use electron cooling in PETRA
\cite{balewski99a,gentner99a,brinkmann99a} so as to
reduce the emittance in HERA-p and to allow for an acceleration
without loss of polarization for most particles in the beam.

\begin{figure}[ht!]
\begin{center}
\begin{minipage}[t]{\columnwidth}
\includegraphics[width=0.49\columnwidth,bb=113 630 336 755,clip]
             {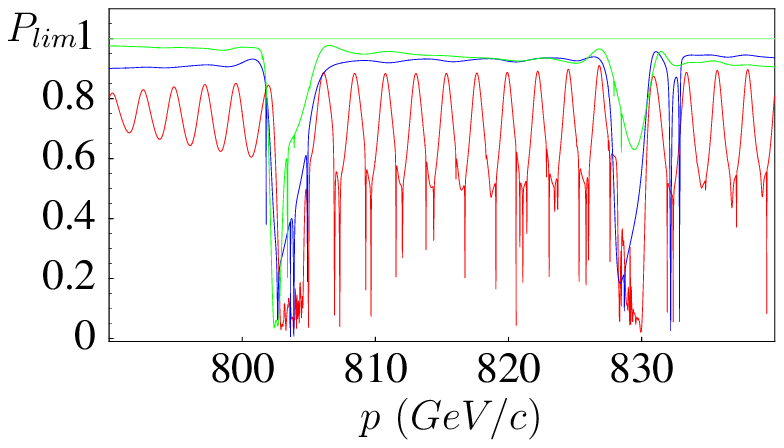}
\includegraphics[width=0.49\columnwidth,bb=113 630 336 755,clip]
             {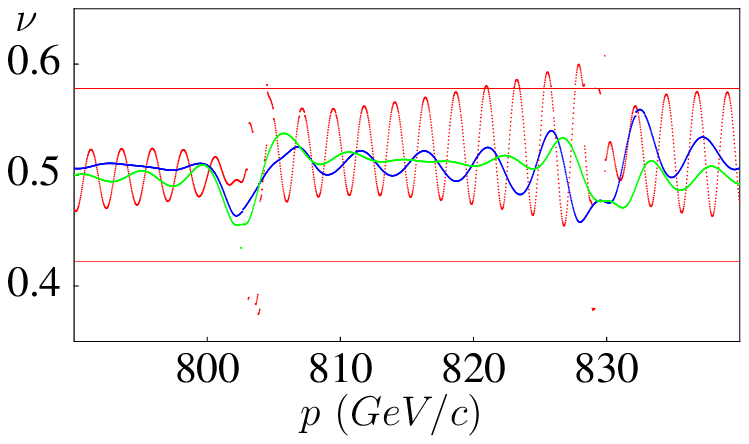}
\includegraphics[width=\columnwidth,bb=136 536 562 771,clip]
             {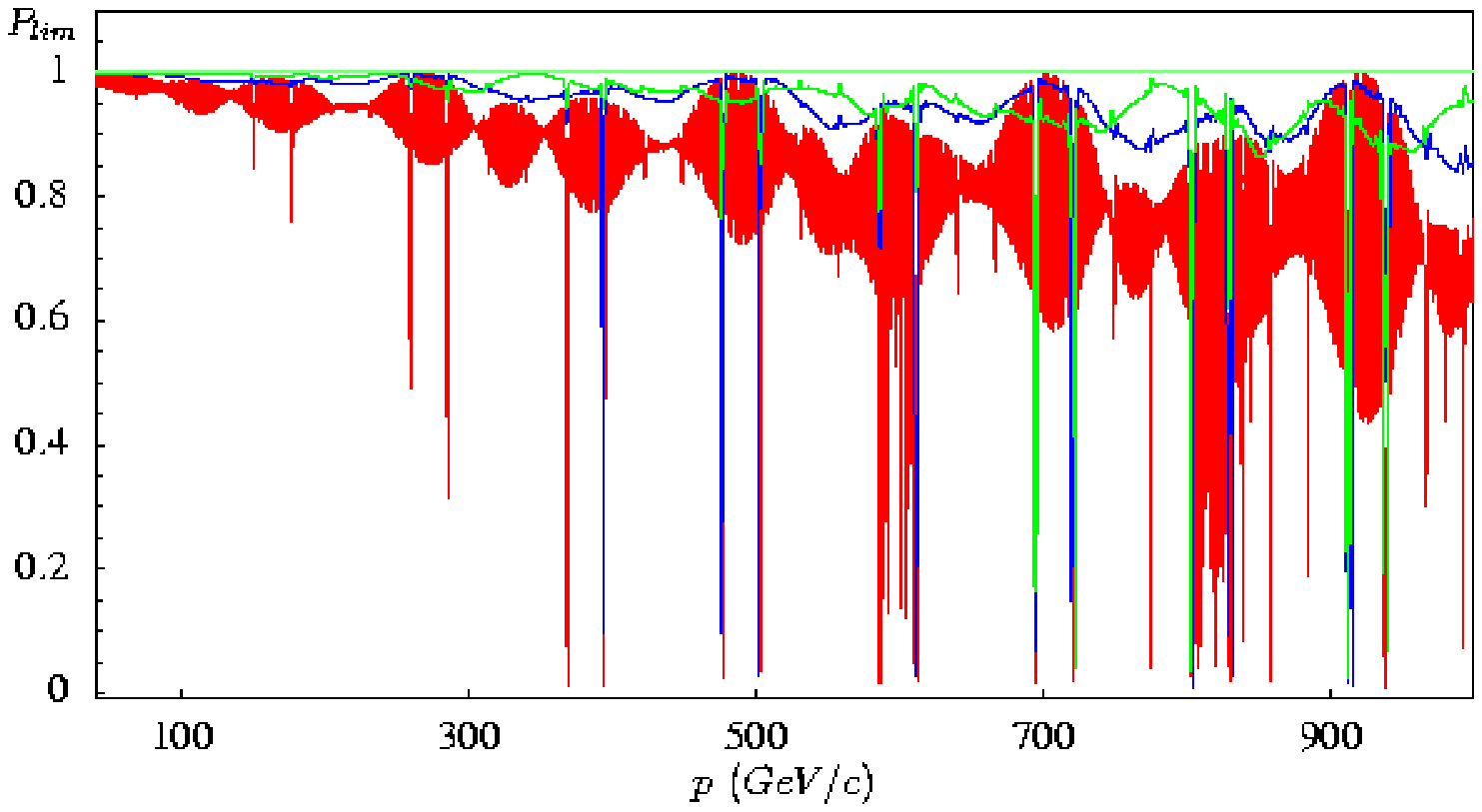}
\includegraphics[width=\columnwidth,bb=136 536 562 771,clip]
             {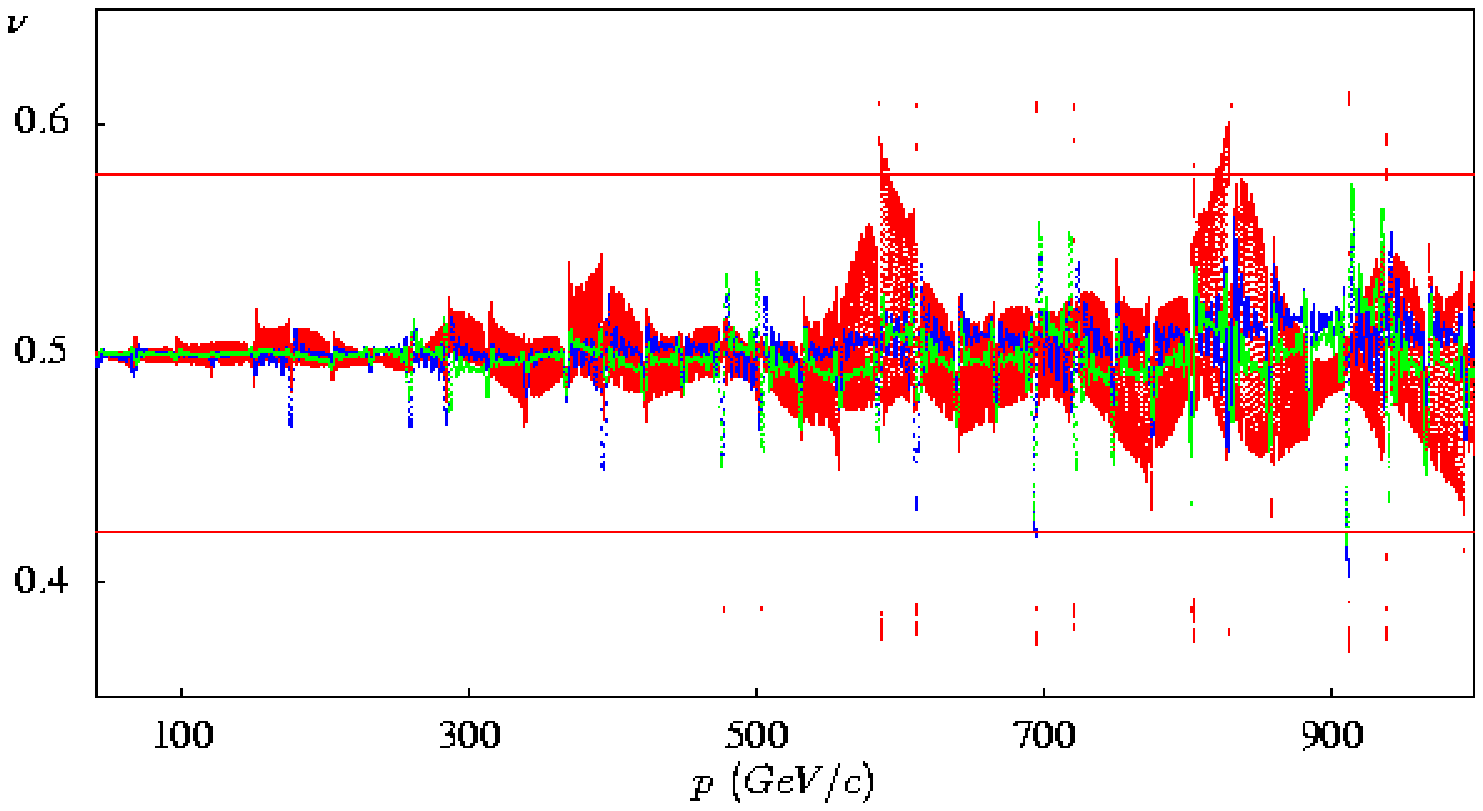}
\end{minipage}
\end{center}
\caption[$P_{lim}$ and $\nu$ with higher-order effects for the snake
matched scheme of 8 Siberian Snakes] {Improvement of the higher-order
$P_{lim}$ and $\nu(J_y)$ by matching 8 snake angles and the orbital
phases.  The snake arrangement is $(\frac{\pi}{2}0000000)o$
(\emph{blue}) and \mbox{$(\frac{\pi}{2}abc0$-$c$-$b$-$a)o$}
(\emph{green}).  As a comparison $P_{lim}$ from SODOM~II is shown for
the snake and phase advance matched four-snake scheme
$0\frac{\pi}{2}\frac{\pi}{2}\frac{\pi}{2})o$ (\emph{red} background
curve). The resonances $\nu=2Q_y$ and $\nu=1-2Q_y$ are indicated (also
\emph{red})
\label{fg:sodom8sn}}
\end{figure}

The excellent performance of the two schemes with 8 Siberian Snakes is
due to much smaller oscillation of the amplitude dependent spin tune
during the acceleration process, and the destructive second order
spin-orbit resonances indicated in Fig.~\ref{fg:sodom8sn}~(bottom and
top-right) are hardly encountered when these snake schemes are chosen.
Correspondingly $P_{lim}$ only drops to small values at very few
energies in Figs.~\ref{fg:sodom8sn}~(center and top-left).

Contemplating all these results, it can be concluded that it is not
possible to give a simple formula for the number of snakes which are
required for a given accelerator since different snake schemes with
the same number of snakes lead to very different stability of spin
motion.  It has even been shown that 8 snakes are not necessarily
better than 4 snakes for the non-flat HERA-p ring
\cite{hoffstaetter99b1}. In the end detailed evaluation is needed.

\section*{Acknowledgments}
\begin{minipage}{\textwidth}
   Desmond Barber's careful reading and improving of the 
   manuscript are thankfully acknowledged.
\end{minipage}

\end{document}